\documentclass{aa}  
\usepackage{natbib}
\usepackage{graphicx}
\usepackage{txfonts}
\usepackage[colorlinks=true,allcolors=cyan]{hyperref}
\begin{document} 
  \title{Populations of filaments from the distribution of galaxies in numerical simulations}
  
  \author{Daniela Gal\'arraga-Espinosa\inst{1}
    \and Nabila Aghanim\inst{1}
    \and Mathieu Langer\inst{1}
    \and C\'eline Gouin\inst{1}
    \and Nicola Malavasi\inst{1}
        }

  \institute{Universit\'e Paris-Saclay, CNRS, Institut d'astrophysique spatiale, 91405, Orsay, France\\
 \email{daniela.galarraga@universite-paris-saclay.fr}
        }

  \date{Received XXX; accepted YYY}

  \abstract{
   {We present a statistical study of the filamentary structures of the cosmic web in the large hydro-dynamical simulations Illustris-TNG, Illustris, and Magneticum at redshift $z=0$. 
   We focus on the radial distribution of the galaxy density around filaments detected using the Discrete Persistent Structure Extractor (DisPerSE). 
   We show that the average profile of filaments presents an excess of galaxy density ($> 5\sigma$) up to radial distances of 27 Mpc from the core.
   The relation between galaxy density and the length of filaments is further investigated showing that short ($L_\mathrm{f} < 9$ Mpc) and long ($L_\mathrm{f} \geq 20$ Mpc) filaments are two statistically different populations. Short filaments are puffier, denser, and more connected to massive objects, whereas long filaments are thinner, less dense, and more connected to less massive structures. These two populations trace different environments and may correspond to bridges of matter between over-dense structures (short filaments), and to cosmic filaments shaping the skeleton of the cosmic web (long filaments). Through Markov Chain Monte Carlo (MCMC) explorations, we find that the density profiles of both short and long filaments can be described by the same empirical models (generalised Navarro, Frenk and White, $\beta$-model, a single and a double power law) with different and distinct sets of parameters.}}

\keywords{(cosmology:) large-scale structure of Universe, methods: statistical, methods: numerical}

\maketitle


\section{Introduction}

On its largest scales, matter in the Universe is not distributed uniformly: it is organised in clusters, filaments, walls, and voids. These structures form the cosmic web \citep{Bond1996}, that is a gigantic network of dark matter and gas, which formed under the effect of gravity from the anisotropic collapse of initial fluctuations of the density field \citep{Zeldovich1970}. The cosmic web was first observed in the Center for Astrophysics (CfA) galaxy survey by \cite{Lapparent1986} and since then has drawn a lot of attention in both observational and simulation fields.

Indeed, the cosmic web, as traced by galaxies, has been observed over the last two decades in many galaxy surveys with increasing resolution and statistics. We note, for example, the Sloan Digital Sky Survey \citep[SDSS,][]{York2000}, the two degree Field Galaxy Redshift Survey \citep[2dFGRS,][]{Colless2003_2dFsurvey}, the Cosmic Evolution Survey \citep[COSMOS,][]{Scoville2007_COSMOSsurvey}, the 6dF Galaxy Survey \citep[6dFGS][]{Jones2009_6dFGSurvey}, the Galaxy and Mass Assembly \citep[GAMA,][]{Driver2011_GAMAsurvey}, and the VIMOS VLT deep survey \citep[VVDS]{Lefevre2005}, VIPERS \citep{Guzzo2014} and SAMI \citep{Bryant2015_SAMIsurvey} surveys.
All of these observations have shown galaxies distributed in filamentary structures, exhibiting the filamentary pattern of the cosmic web.

A deeper understanding of the dynamics of the large scale structures was achieved thanks to N-body simulations \citep[e.g.][]{Bond1996, Springel2005}. Dark matter (DM) was shown to assemble from voids to walls, then flowing to filaments until reaching the highest density regions, the nodes. In addition, N-body simulations provided the means to examine the morphology and density of the different cosmic structures \citep[see e.g.][]{Hahn2007, AragonCalvo2010, Cautin2014}.
More recently, high resolution hydro-dynamical simulations aiming to match some of the current observations (e.g. Illustris \citep{Vogelsberger2014_Illustrissim, Nelson2015Illustris} and IllustrisTNG \citep{Nelson2019}, Magneticum \citep{Hirschmann2014MAGNETICUM, Dolag2015MAGNETICUM, Ragagnin2017MAGNETICUM}, EAGLE \citep{Schaye2015_EAGLEsimu}, or the Horizon-AGN \citep{Dubois2014} simulation) have been performed to allow for a more detailed study and characterisation of the cosmic structures and to follow their evolution in correlation with their tracers, that are baryonic gas and galaxies \citep[e.g.][]{GhellerVazza2016, Martizzi2019a, Gheller2019}.

While a lot of attention has been drawn to the study of clusters in terms of their galaxy content, galaxy properties, gas composition, or density profiles \citep[e.g.][]{Nagai2007, Arnaud2010, Baxter2017, Bartalucci2017, PintosCastro2019, Ghirardini2019}, filaments remain a challenge to characterise.
A hint as to the diversity of the filamentary structures has been shown by some case studies in observational data. For example, \cite{Bonjean2018} studied the pair of clusters A399-A401 with a particular focus on the filament between these two clusters, which turned out to be populated by quiescent galaxies and hot and dense gas. Other studies analysed this structure and detected large-scale accretion shocks thanks to observations in the X-ray \citep{Akamatsu2017} as well as radio emission from the filament \citep{Govoni2019}. Another example from \cite{Malavasi2020_coma} is the analysis of the filamentary environment of the Coma cluster, showing that this structure is part of a network composed by at least three filaments directly connected to Coma.
Moreover, the lack of a standard and unique definition of filaments has led to the development of a variety of algorithms aiming, each with a different approach, to detect and identify the filamentary structures from galaxy distributions or DM particles (e.g. the Discrete Persistent Structure Extractor (DisPerSE) of \cite{Disperse_paper1}, the SpineWeb of \cite{Aragon-Calvo2010_SPINE}, the NEXUS algorithm by \cite{Cautun2013nexus}, the BISOUS model by \cite{Tempel2016bisous}, and more recently the T-Rex method developed by \cite{Bonnaire2019Trex} based on smooth graph structures or the machine learning classification algorithm of \cite{Buncher2019}). Thanks to these detection methods, several statistical studies of galaxies around cosmic filaments have been undertaken \citep[e.g.][]{Malavasi2017, Chen2017_fil_gal, Kuutma2017, Laigle2018, Kraljic2018, Kraljic2019, Bonjean2019filaments, Ganeshaiah2019, Rost2020}. It has been shown that the properties of galaxies strongly depend on their location with respect to filaments. For example, at the core of  filaments, an excess of red  with respect to blue galaxies has been detected \citep{Laigle2016cosmos, Malavasi2017, Kraljic2018}, along with a reduction of the star formation rate (SFR) and an enhancement of stellar masses \citep{Bonjean2019filaments}. 
Similar studies have also been conducted for filamentary structures around clusters to investigate environmental effects on galaxy evolution \citep{Gouin2019, Sarron2019}. 
 In addition, stacking analyses using the Sunyaev-Zel'dovich effect (SZ) have revealed the existence of hot gas in filaments \citep{Tanimura2019, DeGraaff2019, Tanimura2020}, and numerical simulations have helped to characterise this gas \citep[e.g. ][]{Gheller2019, Martizzi2019a}. 

In this paper, we study filaments detected with the DisPerSE algorithm \citep{Disperse_paper1, Sousbie2011b} by means of their galaxy density profiles, in several large numerical simulations, namely Illustris-TNG \citep{Nelson2019}, Illustris \citep{Genel2014ILLUSTRIS, Vogelsberger2014ILLUSTRIS, Nelson2015Illustris} and Magneticum \citep{Hirschmann2014MAGNETICUM, Dolag2015MAGNETICUM, Ragagnin2017MAGNETICUM}. 
We aim to characterise cosmic filaments and identifying differences, if any, in the galaxy density profile of filaments having different lengths and lying in different environments. Inspired by the galaxy cluster literature and following an empirical approach, we also investigate whether some analytical models can be adapted to describe the radial profiles of filaments of the cosmic web.

This paper is organised as follows: in Sect.~\ref{Sect:Data}, we describe the Illustris-TNG, the Illustris and the Magneticum simulations, and the associated  galaxy catalogues used for the identification of the filaments. The filament catalogue and its extraction with DisPerSE are described in Sect.~\ref{Sect}, along with the method for the computation of the galaxy density profiles.
Section \ref{Sect:Results} presents our results in terms of radial galaxy density profiles, filament populations, filament environments and models fitting the density profiles. Finally, we summarise our conclusions in Sect.~\ref{Sect:Conclusions}.

\section{\label{Sect:Data}Data}

We use the outputs of a state-of-the-art large numerical simulation, namely Illustris-TNG\footnote{https://www.tng-project.org} \citep{Nelson2019}, and we compare our findings with the outputs of other large scale simulations, the Illustris \footnote{https://www.illustris-project.org/} simulation \citep{Vogelsberger2014ILLUSTRIS, Genel2014ILLUSTRIS, Nelson2015Illustris} and the Magneticum Pathfinder\footnote{http://www.magneticum.org} \citep{Hirschmann2014MAGNETICUM,Dolag2015MAGNETICUM, Ragagnin2017MAGNETICUM}. We analyse a set of different simulations in order to reach conclusions that are the most possible free from simulation bias.

\subsection{The simulation boxes}

We analyse the gravo-magnetohydrodynamical simulation Illustris-TNG \citep{Nelson2019}. This simulation follows the coupled evolution of DM, gas, stars, and black holes from redshift $z=127$ to $z=0$. Illustris-TNG is run with the moving-mesh code Arepo \citep{Arepo}, and the values of the cosmological parameters are those of the {\it Planck} 2015 \citep{Planck2015Cosmo} results: $\Omega_{\Lambda,0} = 0.6911$, $\Omega_{m,0}=0.3089$, $\Omega_{b,0}=0.0486$, $\sigma_{8}=0.8159$, $n_s=0.9667$ and $h=0.6774$. We focus on the simulation box TNG300-1 at a redshift $z=0$. It consists in a cube of around 302 Mpc side length with a DM resolution of ($m_{\mathrm{DM}} = 4.0 \times 10^{7} \mathrm{M_{\odot}}/h$ and $N_{\mathrm{DM}} =2500^{3}$).
We choose to use the largest simulation volume with the highest mass resolution to accurately describe cosmic filaments down to small scales. Throughout this paper this simulation is referred to as the reference.

In Sect.~\ref{Subsect:radial_density_profile} and \ref{SubSect:bybinsL}, other runs of the Illustris-TNG series are analysed for specific purposes. The TNG300-2 simulation box is used to test for the effects of resolution. This box is the medium resolution run of the TNG300 series, meaning that it has the same characteristics as the TNG300-1 simulation except for a mass resolution of $m_{\mathrm{DM}} = 3.2 \times 10^{8} \mathrm{M_{\odot}}/h$, that is eight times lower than the reference.
The outputs of two other boxes, namely TNG100-2 and Illustris-2, are also studied in order to asses the impact of the simulations' intrinsic physical models (of star formation, cooling, feedback, etc.) on our results. Indeed, these two boxes have the same initial conditions, similar box sizes and mass resolutions (which enables direct comparison, see Table \ref{Table:sims_overview} for details), but they involve significantly different physical models. The Illustris-TNG project is the successor of the original Illustris simulation \citep{Nelson2015Illustris, Genel2014ILLUSTRIS, Vogelsberger2014ILLUSTRIS}, and was specifically calibrated on the observed galaxy properties and statistics \citep{Nelson2019}. In this sense, the main differences between the TNG and the Illustris physical models reside in the formation, growth and feedback of supermassive black holes, as well as in the galactic winds, stellar evolution and gas chemical enrichment models. A full and detailed comparison of the two can be found in \cite{Pillepich2018TNGmodel}.\\

Finally, we use the outputs of the Magneticum simulation \citep{Hirschmann2014MAGNETICUM, Dolag2015MAGNETICUM, Ragagnin2017MAGNETICUM} at the smallest redshift available, $z=0.066$. It consists in a cube of 500 Mpc side length with a DM resolution of $m_{\mathrm{DM}} = 6.9 \times 10^{8} \mathrm{M_{\odot}}/h$. The cosmological parameters of this simulation are different from Illustris-TNG, as the cosmology is described according to results of  the seven-year Wilkinson Microwave Anisotropy Probe (WMAP7) data \citep{Komatsu2011WMAP7}:
$\Omega_{\Lambda,0} = 0.728$, $\Omega_{m,0}=0.272$, $\Omega_{b,0}=0.0456$, $\sigma_{8}=0.809$, $n_s=0.963$ and $h=0.704$.

An overview of the main characteristics of the simulations used in this work can be found in Table~\ref{Table:sims_overview}.

\begin{table*}
\caption{Overview of the five simulations and of the derived filament catalogues used in this work. The TNG300-1 simulation \citep{Nelson2019} is used as reference for the analysis, and we compare to it the results obtained with Magneticum \citep{Hirschmann2014MAGNETICUM, Dolag2015MAGNETICUM}. TNG300-2 is compared to the reference TNG300-1 to asses the effect of resolution. TNG100-2 and Illustris-2 simulations \citep{Genel2014ILLUSTRIS, Vogelsberger2014ILLUSTRIS, Nelson2015Illustris} are compared to each other to study the effects of the intrinsic physical models on our results.}
\label{Table:sims_overview}     
\centering  
\begin{tabular}{ c  c  c  c  c  c }
 \hline\hline   
 \textbf  & \textbf{TNG300-1} & Magneticum & TNG300-2 & TNG100-2 & Illustris-2 \\ \hline
 
   Box size [$\mathrm{Mpc}^3$] &  $302.6^3$ & $500^3$ & $302.6^3$ & $110.7^3$ & $ 106.5^3$ \\
   
   DM resolution [$\mathrm{M_{\odot}}/h$] & $4.0 \times 10^{7}$ & $6.9 \times 10^{8}$ & $3.2 \times 10^{8}$ & $4.0 \times 10^{7}$ & $3.5 \times 10^{7}$ \\
      
   Cosmology & \textit{Planck} 2015 & WMAP7 & \textit{Planck} 2015 & \textit{Planck} 2015 & WMAP7 \\
   
    Analysed volume [$\mathrm{Mpc}^3$] &  $250^3$ & $450^3$ & $250^3$ & $90^3$ & $90^3$ \\
   
    Density of tracers [$\mathrm{Mpc}^{-3}$]& $10.0 \times 10^{-3}$ & $10.1 \times 10^{-3}$ & $5.4 \times 10^{-3}$ & $10.1 \times 10^{-3}$ & $16.9 \times 10^{-3}$ \\

    \\ 
    Number of filaments & 5550 & 38,278 & 2885 & 213 & 223 \\
   Min and Max filament lengths [Mpc] & [0.4, 65.6] & [0.1, 93.6] & [0.6, 85.2] & [1.5, 54.0] & [1.4, 52.1] \\
   Mean filament length [Mpc] & 10.9 & 12.2 & 14.3 & 10.7 & 11.2 \\
   Median filament length [Mpc] & 8.8 & 9.8 & 11.4 & 8.3 & 9.2 \\
   \\
    Comments & \textbf{Reference} & Comparison & Resolution effects & Physical models & Physical models \\
  \hline
 \end{tabular}
\end{table*}

\subsection{The galaxy catalogues}\label{SubSect:subhalos}

In the following we describe the procedure we followed to build the galaxy catalogues used in this work. For the Illustris/Illustris-TNG simulations we show only the example of TNG300-1, as the galaxy catalogues of the other runs are built in the exact same way.

For the reference simulation (TNG300-1) and for the other simulations of the Illustris/Illustris-TNG series, we build the catalogues of galaxies from the corresponding SUBFIND subhalo catalogues, which are lists of gravitationally bound overdensities. The SUBFIND algorithm was first introduced in \citet{halosandsubhalos} for DM simulations, and  adapted by \citet{subhalos2} to treat gas and star particles of hydrodynamical simulations. 

From the original subhalo catalogue and following \citet{Nelson2019}, we discard all the objects that have been flagged by Illustris-TNG as subhalos but  have not followed the usual processes of galaxy formation (field \texttt{SubhaloFlag} in the catalogue). This avoids  contamination by fragments or clumps originated from baryonic processes. These objects account for a number of $50,098$ subhalos and represent a very small fraction ($\sim 0.03 \%$) of the initial catalogue. 

\begin{figure}[h!]
   \centering
   \includegraphics[width=0.5\textwidth]{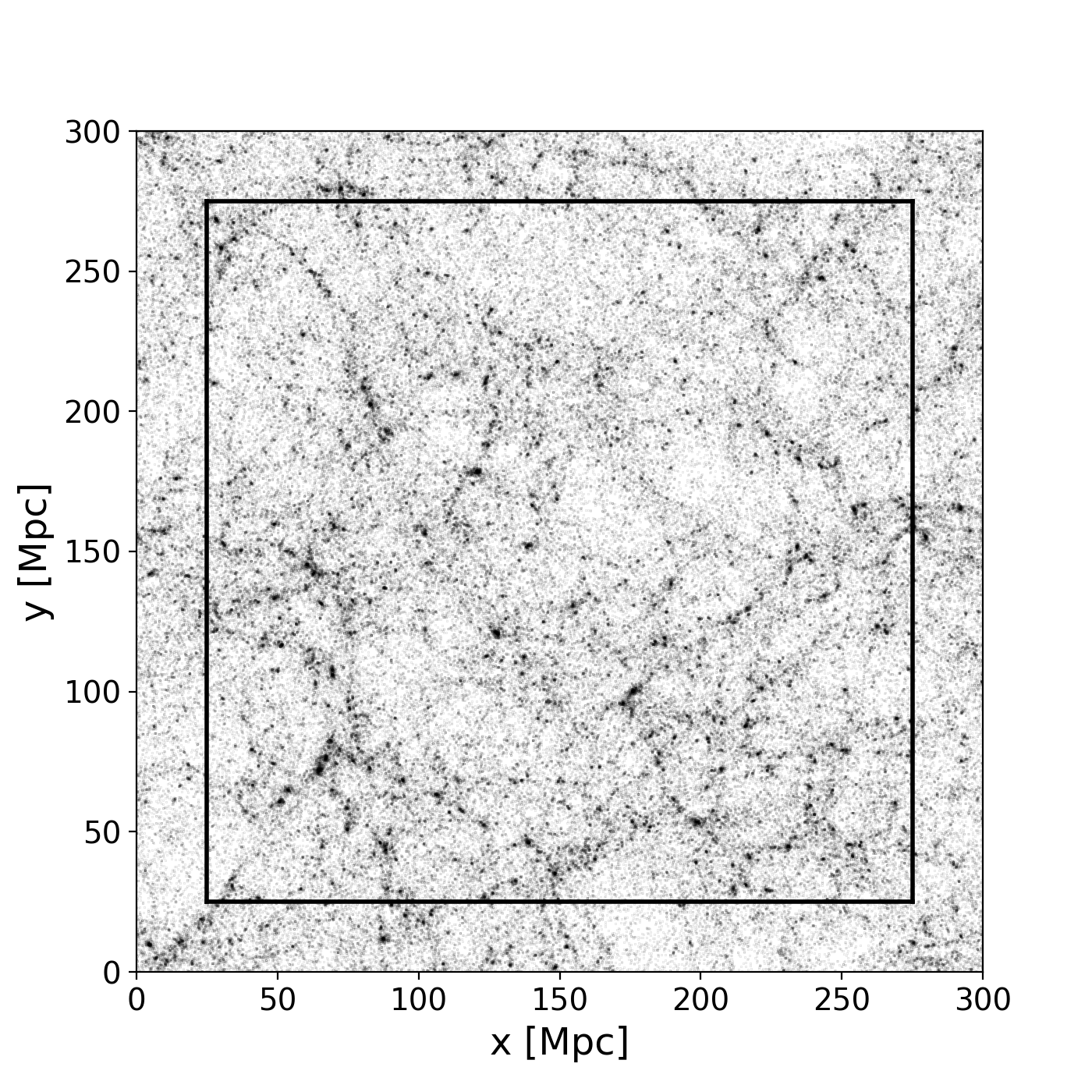}
   \caption{Galaxies (subhalos) of the TNG300-1 simulation box after  applying the selection described in Sect.~\ref{SubSect:subhalos}. These galaxies have stellar masses in the range $10^{9} \le M_{*} [\mathrm{M_{\odot}}] \le 10^{12}$ and constitute our reference catalogue. The full 3D box has been projected onto the $xy$ plane. The black square represents the boundaries of the central volume of $[250 \, \mathrm{Mpc}]^3$ considered for the analysis (see Sect.~\ref{SubSec:filament_catalogue}).}
    \label{Fig:gal}
    \end{figure}

We apply a cut in stellar mass to the remaining subhalos by using the field \texttt{SubhaloMassType} of the SUBFIND catalogue. Following the observational limits of \citet{Brinchmann2004} and  \citet{Taylor2011}, we require the stellar mass of the subhalos to lie in the range $10^{9} \le \mathrm{M}_{*} [\mathrm{M_{\odot}}] \le 10^{12}$. After this selection in mass, $275,818$ subhalos remain. Figure \ref{Fig:gal} presents a projection in the $xy$ plane of the distribution of these subhalos in which the filamentary structure of the cosmic web is already discernible.
From now on, these objects are referred to as galaxies and constitute our TNG300-1 galaxy catalogue.\\

Throughout this paper, the results found using the TNG300-1 galaxies serve as our reference, to which we compare the results obtained with the large \texttt{Box2/hr} galaxy catalogue of the Magneticum simulation. We applied the same stellar mass cut to this simulation and we ended up with $1,266,306$ galaxies in the mass range $10^{9} \le \mathrm{M}_{*} [\mathrm{M_{\odot}}] \le 10^{12}$. The Magneticum galaxy catalogue is statistically roughly $4.6$ times larger than the reference catalogue, which is not surprising considering that the simulation box is $2.5$ times bigger.

\section{\label{Sect}Methods}

\subsection{\label{SubSec:Disperse}Filament detection}

We detect and extract the skeleton of the cosmic web using the publicly available algorithm Discrete Persistent Structure Extractor \citep[DisPerSE,][]{Disperse_paper1, Sousbie2011b} applied to the galaxy catalogues described in the previous section.

DisPerSE is an algorithm that detects filaments based on the topology of the density field, which is computed from the distribution of input particles (galaxies, in this case), using the Delaunay Tessellation Field Estimator \citep[DTFE,][]{SchaapWeygaert2000, WeygaertSchaap2009}. 
In this work as a first step, we smoothed once the Delaunay density field by averaging the value of the density at each vertex (which corresponds to the position of galaxies) with the surrounding vertices of the tessellation. This smoothing was performed in order to minimise the contamination by shot noise and to prevent the identification of small scale, possibly local or spurious features \citep{Malavasi2020_sdss, Malavasi2020_coma}.

In order to find the filamentary structures, the algorithm first identifies the critical points of the density field using Discrete Morse theory. These points are located where the gradient of the field vanishes, meaning that the critical point is either a maximum, a minimum or a saddle point of the density field. Filaments are then defined as ridges of the Delaunay density field: each filament is a set of segments connecting a maximum to a saddle point. 
The user can choose the significance of the detected filaments by fixing the persistence ratio of the corresponding pair of critical points. In this work we extracted the 3D skeleton with a $3\sigma$ persistence level. The choice of this value was motivated by the fact that a lower persistence value resulted in a larger number of small scale features likely to emerge from the noise, whereas a higher value provoked a significant drop in the number of filaments as only the most reliably identified structures were kept. Furthermore, the $3\sigma$ persistence skeleton gave visually the best coincidence of positions of the maximum density critical points with respect to the over-dense regions of the input set of galaxies.

We applied a final smoothing to the extracted skeleton by shifting each segment extremity to the middle position of the contiguous extremities.
This smoothing was done with the aim of alleviating the effect of shot noise on the geometry of the filaments, which would otherwise generate sharp and possibly nonphysical edges between the segments composing the filaments.

\begin{figure}
   \centering
   \includegraphics[width=0.5\textwidth]{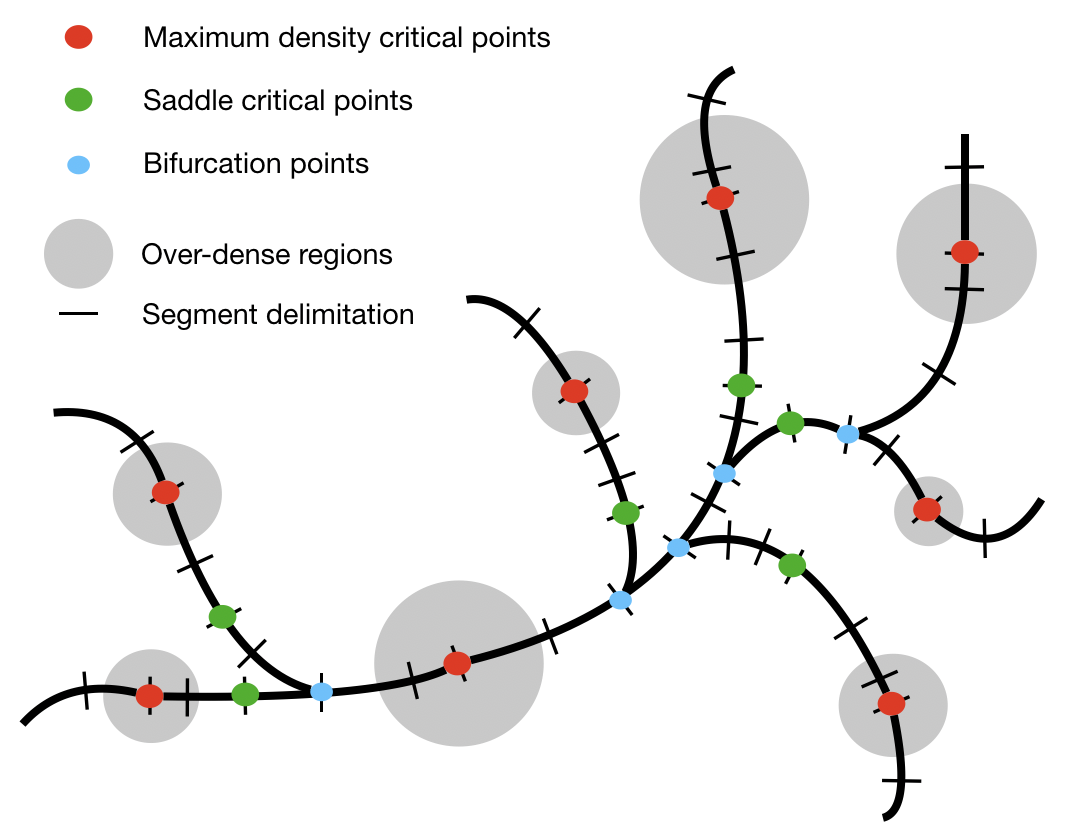}
   \caption{Illustration of the 3D DisPerSE skeleton projected on a 2D plane. 
   Segments of filaments (thick black lines) are organised into portions delimited by critical points (in red, green and blue respectively for maxima, saddles and bifurcations). Maximum density critical points are located in over-dense regions (represented by grey discs) identified from the Delaunay density field.
   This illustration shows a total of 44 segments, organized in 18 portions forming 13 filaments, defined as the sets of segments connecting maxima to saddles.}
    \label{Fig:schema_skel}%
    \end{figure}

A 2D illustration of the extracted DisPerSE skeleton is shown in Fig.~\ref{Fig:schema_skel}. In this figure, we see that some filaments share common segments that should not be counted twice in the statistical analysis.
Therefore, we applied the DisPerSE `breakdown' procedure to break filaments sharing one or more segments into separate portions of filaments. This is done by introducing bifurcation points at the positions where  filaments merge. In this way, we obtain portions of filaments delimited either by maxima, saddle or bifurcation points.

Finally, we noticed an excess of critical points near the edges of the simulation box. These can be removed using dedicated flags in DisPerSE \citep{Disperse_paper1}. However, in a more conservative way, we chose to completely disregard the regions of thickness 25 Mpc at the edges of the TNG300-1 simulation box. This value was chosen by visual inspection and corresponds to twice the width of the zones presenting this excess. We thus kept $161,357$ galaxies inside the central volume of $[250 \, \mathrm{Mpc}]^3$, delimited by the black square in Fig.~\ref{Fig:gal}.

For the sake of comparison, we apply exactly the same procedure described in the previous paragraphs to the Magneticum, TNG300-2, TNG100-2 and Illustris-2 simulations to detect and extract the filaments from their corresponding galaxy catalogues. In theses cases the border effects are also corrected by disregarding the regions of 25 Mpc thickness at the borders of the largest boxes (i.e. Magneticum and TNG300-2), and we limited this thickness to 10 Mpc in the smallest boxes (i.e. TNG100-2 and Illustris-2), for the sake of statistics.

\subsection{\label{SubSec:filament_catalogue}Filament catalogues}

The filament catalogue of the TNG300-1 simulation is built by reconnecting the portions lying inside the sub-box of volume $[250 \, \mathrm{Mpc}]^3$ (see Sect.~\ref{SubSec:Disperse}) to obtain filaments linking a maximum density critical point to a saddle point. This topological definition is physically motivated by the fact that matter within a filament is presumed to flow from the saddle (a local minimum in the axis of a filament but a maximum in the transverse axis) to the point of maximum density \citep{Kraljic2019}.
For example in the illustration of Fig.~\ref{Fig:schema_skel}, we count 13 reconstructed filaments (from red to green points), 18 portions (connecting any type of critical point) and 44 segments. We thus retrieve, in the $[250 \, \mathrm{Mpc}]^3$ central box of the TNG300-1 simulation, a total of $5550$ maximum-saddle filaments composed by $25,818$ different segments (themselves organized in $6920$ portions). The same procedure of filament reconstruction is applied to the Magneticum, TNG300-2, TNG100-2 and Illustris-2 simulations in their corresponding sub-boxes of respective volumes $[450 \, \mathrm{Mpc}]^3$, $[250 \, \mathrm{Mpc}]^3$, $[100 \, \mathrm{Mpc}]^3$ and $[100 \, \mathrm{Mpc}]^3$.

The number of filaments retrieved in each simulation is shown in Table~\ref{Table:sims_overview}. We note that this number depends on: \textit{(i)} the size of the simulation box, and \textit{(ii)} the density of tracers, $n_\mathrm{gal}^\mathrm{box}$, corresponding to the number of galaxies per $\mathrm{Mpc}^3$ (see the fifth line of Table~\ref{Table:sims_overview}). The first item explains the fact that, at fixed density of tracers, smaller simulation boxes contain less filaments than bigger volumes, as expected by statistical considerations. The second item justifies the reduced number of filaments in TNG300-2 with respect to the reference TNG300-1. Indeed, the lower resolution of TNG300-2 results in less refined simulated features in the same volume than the reference, and consequently the density of tracers is significantly lower (by a factor of $\sim 2$) in TNG300-2 than in its higher resolution counterpart, leading to less filaments.

\begin{figure}
   \centering
   \includegraphics[width=0.5\textwidth]{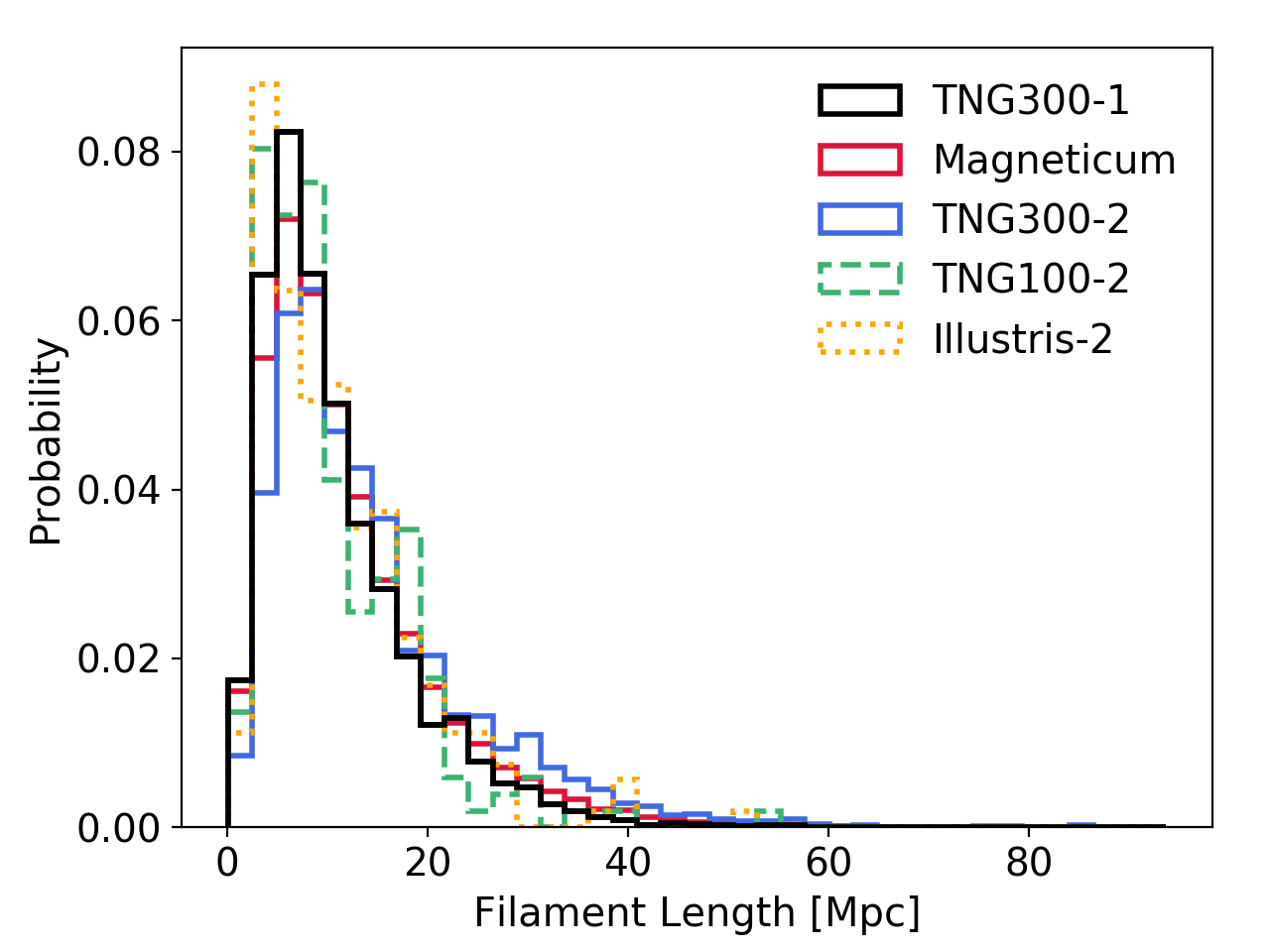}
   \caption{Length distribution of filaments of the five different simulations used in this work (see Table~\ref{Table:sims_overview}). By definition, filaments are sets of contiguous segments connecting maximum density critical points to saddles (see Sect.~\ref{SubSec:filament_catalogue}).}
    \label{Fig:lengths_distribution}%
    \end{figure}

We compute the length of each filament as the sum of the lengths of all the segments belonging to it, from maximum to saddle. 
The length distributions of the filaments found in the TNG300-1, Magneticum, TNG300-2, TNG100-2 and Illustris-2 catalogues are shown in Fig.~\ref{Fig:lengths_distribution}, respectively in black, red, blue, green and yellow. These distributions have an exponential tail, in agreement with previous \citep{Bond2010b, Choi2010} and more recent \citep{Malavasi2020_sdss} findings. This distribution hints at the hierarchical and multi-scale character of the filamentary network with significantly more short (peak at $\sim 5-6$ Mpc for TNG300-1) than long filaments \citep[e.g.][]{Cautin2014}.
The maximum, minimum, mean and median filament lengths of these distributions are also presented in Table~\ref{Table:sims_overview}.\\
Apart from reducing the number of filaments, a low density of tracers (i.e. less galaxies in the same volume) also leads to a skeleton with fewer short filaments and to filaments that are longer on average. Thus the peak of the distribution is shifted towards higher values. We clearly see this shift in the TNG300-2 length distribution of Fig.~\ref{Fig:lengths_distribution} (blue histogram). Following the same argument, we expect the Illustris-2 peak to be shifted towards shorter lengths because this catalogue has the highest value of $n_\mathrm{gal}^\mathrm{box}$ among our set of simulations. This effect is seen in Fig.~\ref{Fig:lengths_distribution} (the yellow histogram peaks at $L_\mathrm{f} \sim 3.7$ Mpc) despite the statistical limitations due to the very reduced number of filaments (223) detected in the Illustris-2 box.

\subsection{\label{SubSect_radial_density_profiles_METHOD}Galaxy density around filaments}

We compute the radial profiles of galaxy density around filaments, that is their density along the perpendicular direction to the filament spine (hereafter called $r$). In practice, the recovered radial profile is the average of the galaxy density distribution around individual segments.  

\begin{figure}
   \centering
   \includegraphics[width=0.4\textwidth]{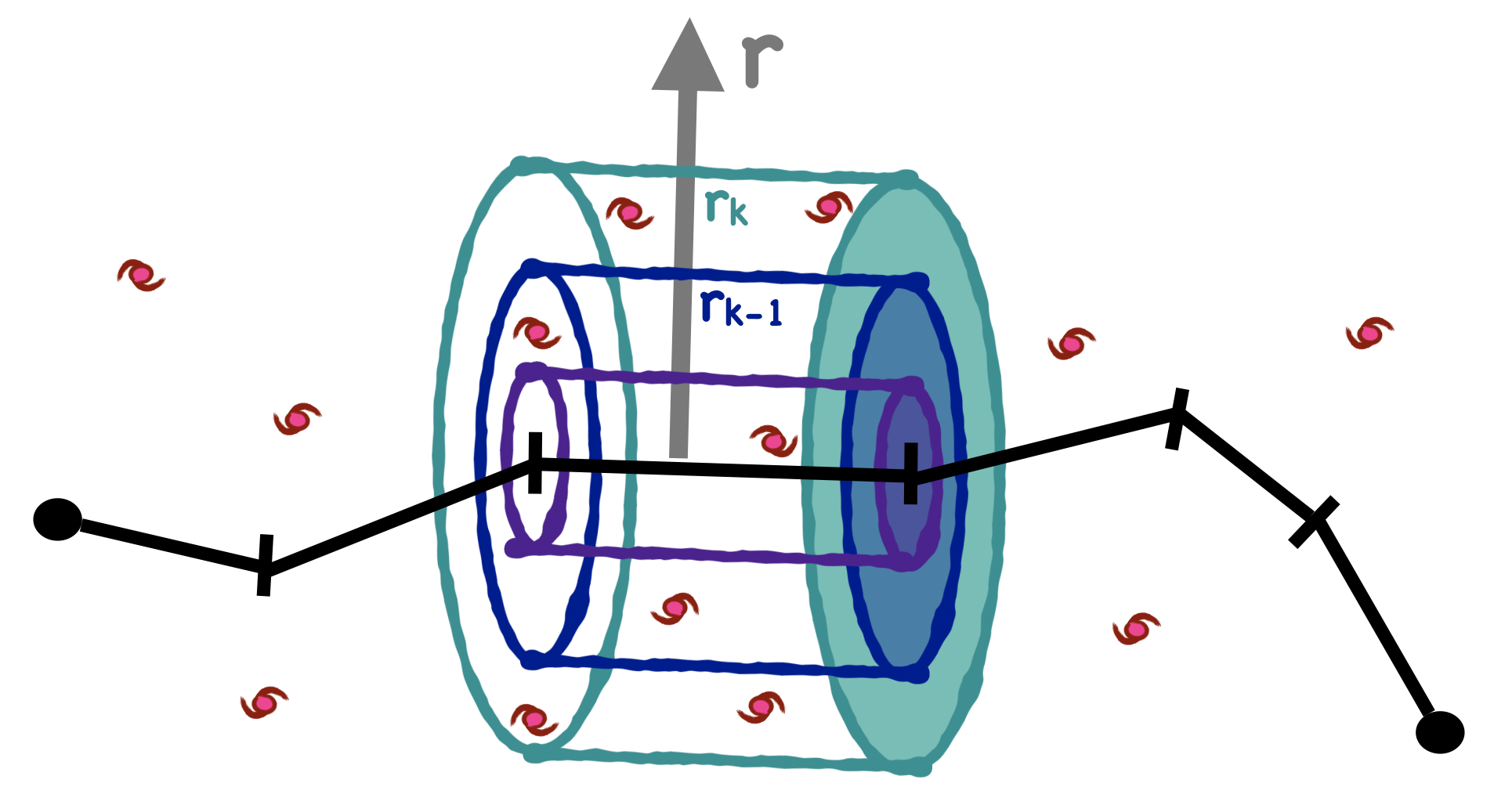}
   \caption{Illustration of the method to compute radial density profiles of filaments (see Sect.~\ref{SubSect_radial_density_profiles_METHOD}). The radial distance to the axis of the filament is called $r$.}
    \label{Fig:density_method}%
    \end{figure}

We start by estimating the radial profile around a segment $i$ of a filament by counting the number of galaxies in concentric cylindrical shells around the axis of the segment, as illustrated in Fig.~\ref{Fig:density_method}, according to the equation:
    \begin{equation}
        n^i(r_k) = \frac{N_k}{\pi (r^2_{k} - r^2_{k-1}) \, l_i}.
    \end{equation}
Here $l_i$ is the length of  segment $i$, the index $k$ stands for the cylindrical shell of thickness $r_{k} - r_{k-1}$ and $N_k$ is the number of galaxies in the $k$-th shell. We bin the radial distance $r$ in $20$ equally spaced logarithmic bins, starting from the core of the segment up to $r = 100$ Mpc. 

The radial profile of galaxy density around filaments is then computed by averaging the profiles of all the segments. In this average, each segment is counted only once, regardless of the number of  reconstructed filaments (see Sect.~\ref{SubSec:filament_catalogue}) to which it belongs.

Errors on the filament profiles are computed by bootstrapping over the set of individual segment profiles, i.e. for a set of $N$ segments, we randomly select (with replacement) $N$ profiles which we average together. We repeat this procedure $N$ times in order to have $N$ averages. The errors of the filament profile are thus the dispersion of these $N$ averages.\\

In order to retrieve the properties that are solely specific to filaments, we remove from the analysis the contribution of the nodes. These over-dense structures are determined by the position of the maximum density critical points of the Delaunay density field (hereafter called CPmax). For simplicity, we consider nodes as spherically symmetric objects and assume that the CPmax lie in their centre. Therefore, we compute the radius $R_{200}$ associated with each CPmax from the DM particles of the TNG300-1 simulation and we mask the galaxies and filament segments inside spheres of radius $3 \times R_{200}$ centred on the positions of the CPmax. The tests and details concerning the choice of mask are discussed in Appendix \ref{Appendix:masks}. In a similar way, the nodes of the Magneticum, TNG300-2, TNG100-2 and Illustris-2 simulations are removed by applying a $3 \times R_{200}$ mask to the positions of the CPmax. Notice that for the Magneticum simulation the $R_{200}$ radii were computed from the galaxy distribution instead of the DM particles. 
In the following, the masked galaxies and segments are removed from the analysis.\\

\begin{figure}
   \centering
   \includegraphics[width=0.25\textwidth]{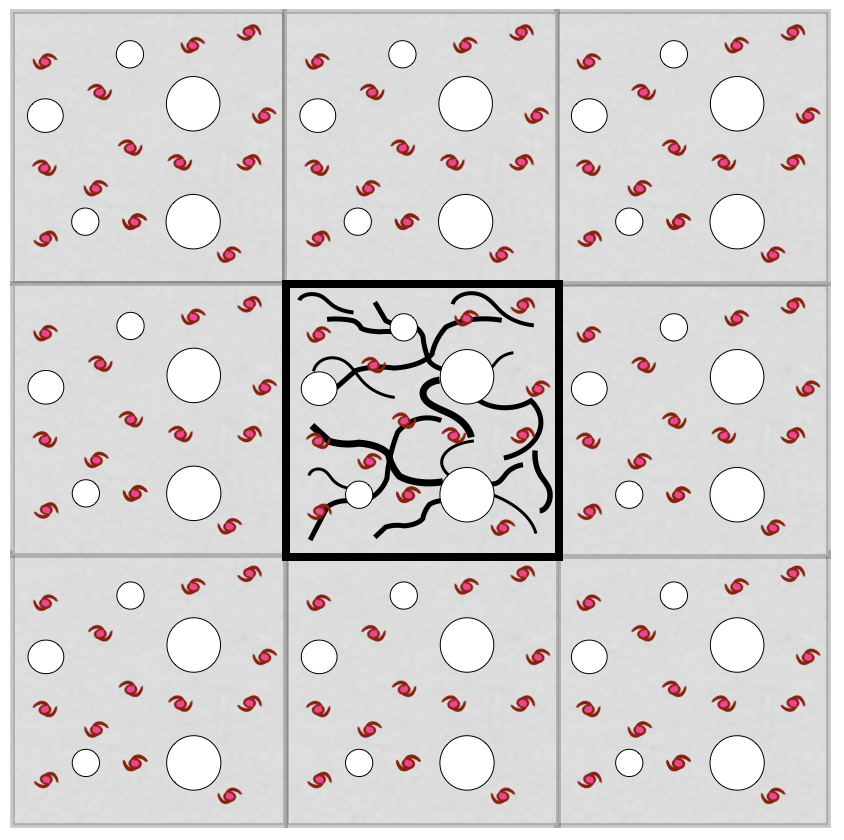}
   \caption{Illustration in 2D of the central volume and the replicated boxes. The galaxy distribution is replicated at the borders of the central volume where the filaments are defined. This is done with the aim of reducing the effects of a volume-limited box in the density profiles. The white circles correspond to the masked regions (see Sect.~\ref{SubSect_radial_density_profiles_METHOD}).}
    \label{Fig:schema_replication}%
    \end{figure}

Finally, we aim at reducing the effects of a volume-limited box in the density profiles. We thus replicate the galaxy distributions at the borders of the central volume of the TNG300-1 simulation in order to construct a $[3 \times 250 \, \mathrm{Mpc}]^3$ volume. The filaments, defined in the central volume of $[250 \, \mathrm{Mpc}]^3$ are not replicated, as is shown in the 2D illustration of Fig.~\ref{Fig:schema_replication}. As we focus only on statistical trends of the galaxy density field, this concatenation (which only affects the continuity of the field at the borders) does not undermine our results. The same replication procedure is applied to the Magneticum, TNG300-2, TNG100-2 and Illustris-2 simulations to construct $[3 \times 450 \, \mathrm{Mpc}]^3$, $[3 \times 250 \, \mathrm{Mpc}]^3$, $[3 \times 90 \, \mathrm{Mpc}]^3$ and $[3 \times 90 \, \mathrm{Mpc}]^3$ volumes, respectively.\\

\section{\label{Sect:Results}Results}


\subsection{\label{Subsect:radial_density_profile}Average density profile of galaxies around filaments}

   \begin{figure}
   \centering
   \includegraphics[width=0.53\textwidth]{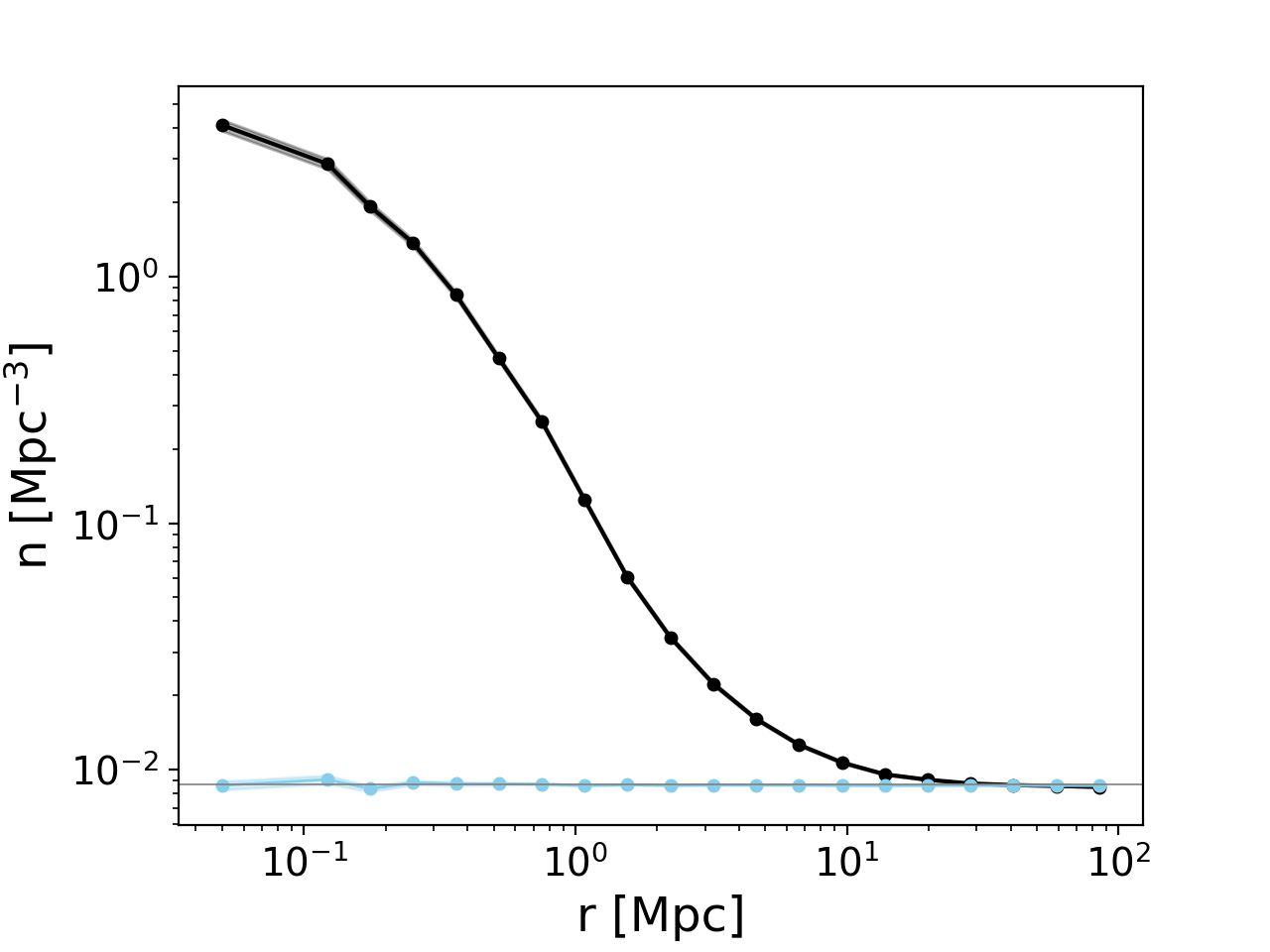}
   \includegraphics[width=0.53\textwidth]{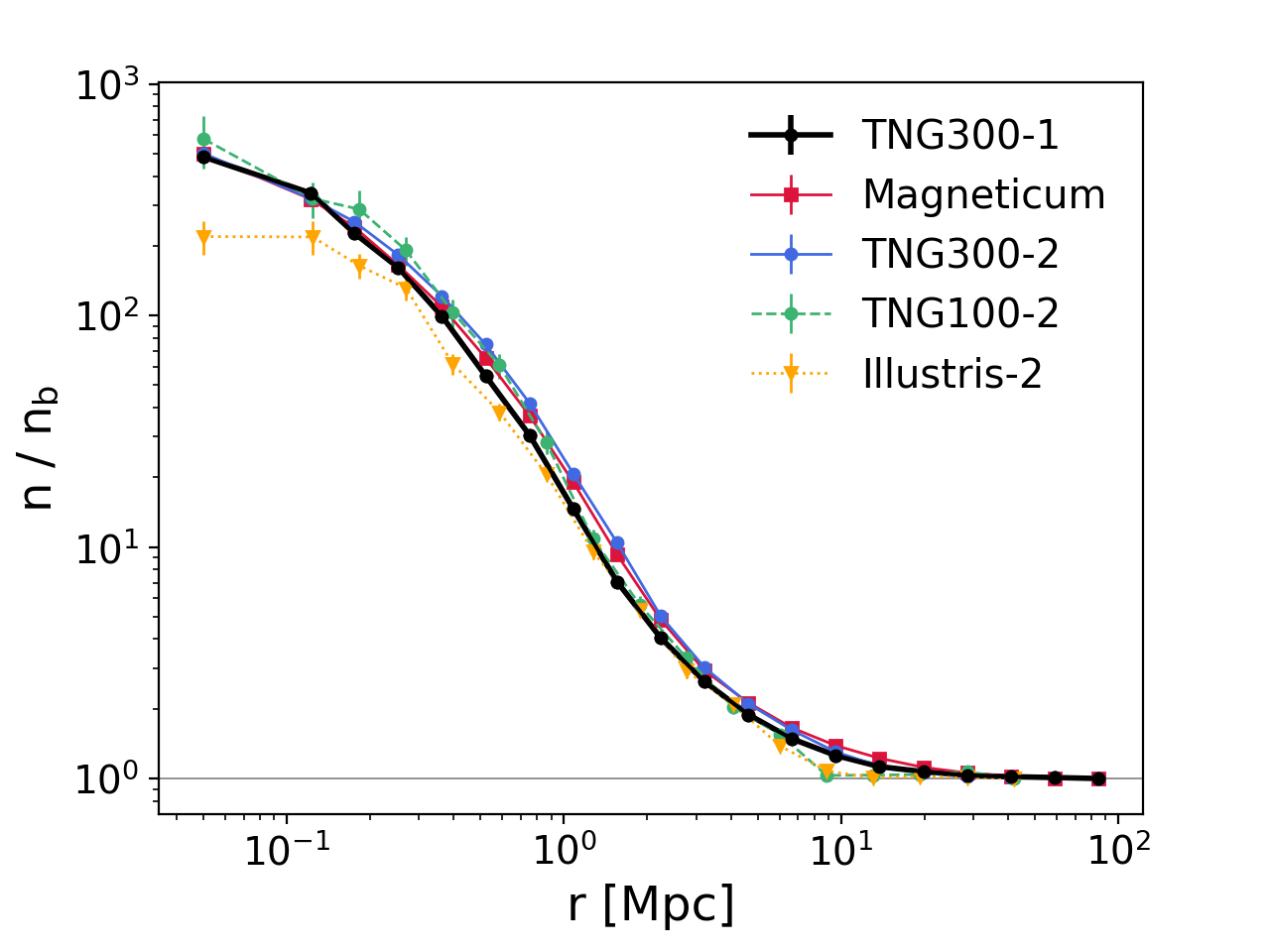}
   \caption{\textit{Top:} Radial density profile of galaxies around filaments of the reference catalogue (TNG300-1). The black curve shows the average galaxy density of all filaments. The blue curve presents the null-test, which is obtained by computing the density of galaxies around randomly shifted filament positions (see Sect.~\ref{Subsect:radial_density_profile}). The gray horizontal line represents the mean galaxy density of the simulation box. \textit{Bottom:} Radial density profile of filaments of the TNG300-1 (black curve), Magneticum (red), TNG300-2 (blue), TNG100-2 (green) and Illustris-2 catalogues (yellow curve). For the sake of comparison, the profiles have been rescaled by their background densities.}
    \label{Fig:profile3_nulltest}
    \end{figure}
    
As explained in Sect.~\ref{SubSect_radial_density_profiles_METHOD}, the density profile of a filament is the average profile of its segments. Hence, the density profile of all the filaments is computed by averaging all the segment profiles in the catalogue. We show the resulting average profile for the TNG300-1 filaments in the upper panel of Fig.~\ref{Fig:profile3_nulltest} (black curve). We also show (blue curve) the corresponding null-test. The latter is computed by shifting the positions of all the filaments of the catalogue by random offsets and computing the galaxy density profile of these shifted filaments as described in Sect.~\ref{SubSect_radial_density_profiles_METHOD}. This procedure is repeated 100 times to create 100 different realisations of the original catalogue of 5550 filaments. The values of the null-test are given by the average profiles of these $5550 \times 100$ randomly shifted filaments.

   \begin{figure*}[h]
   \centering
   \includegraphics[width=1\textwidth]{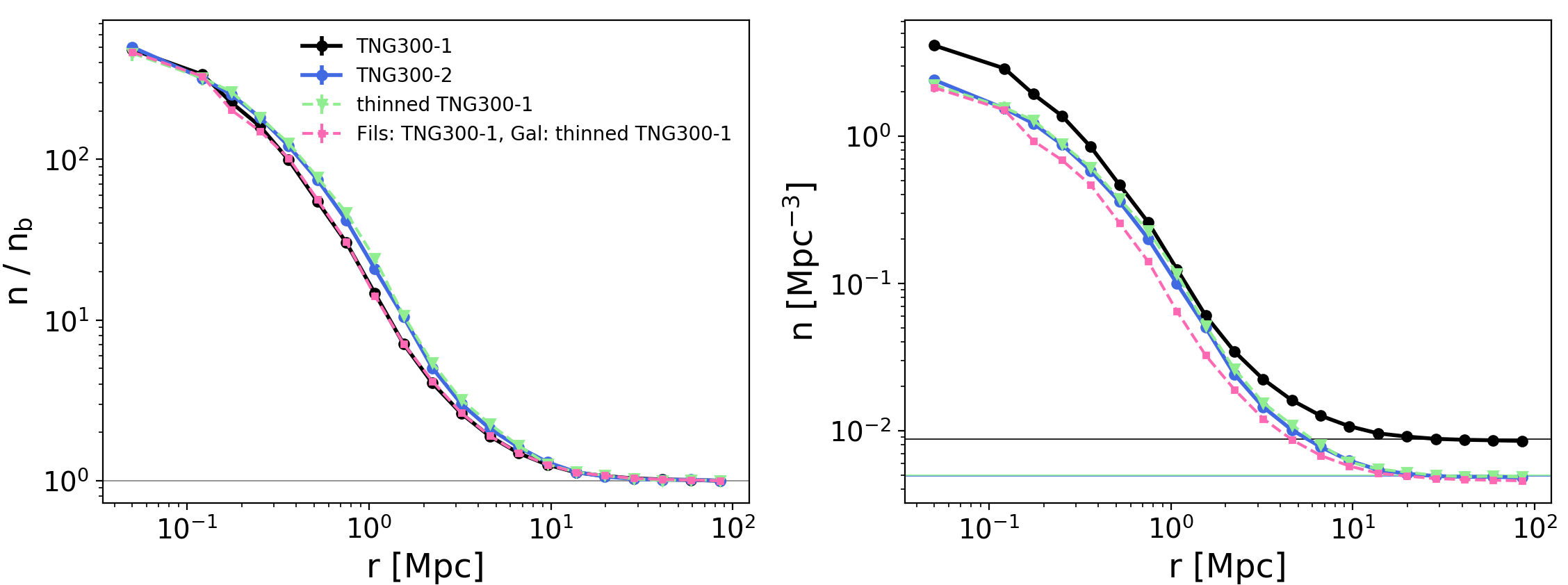}
   \caption{Study of resolution effects in the density profiles (see Sect.~\ref{Subsect:radial_density_profile}). The black and blue curves present the galaxy density profiles of the TNG300-1 and TNG300-2 simulations, as presented in Fig.~\ref{Fig:profile3_nulltest}. The green dashed curve with triangles corresponds to the density profiles of the thinned TNG300-1 galaxy catalogue around and the corresponding DisPerSE filaments (see text in Sect.~\ref{Subsect:radial_density_profile}). The pink dashed profile with squares represents the density of the thinned TNG300-1 galaxy catalogue around the reference TNG300-1 original skeleton. \textit{Left:} Rescaled density profiles. \textit{Right:} Density profiles at their respective galaxy backgrounds.}
    \label{Fig:resolution}
    \end{figure*}
    
The average galaxy density profile of filaments exhibits a clear excess with respect to the background density. The significance at the core  of the filaments with respect to the null-test is found to be $\sim 20 \, \sigma$. We note that the significance of the detection is above $5 \sigma$ up to radial distances of $r = 27$ Mpc. The galaxy density decreases with increasing distance $r$, until it reaches the average density of the simulation box (gray horizontal line). The null-test of Fig.~\ref{Fig:profile3_nulltest} yields the average galaxy density of the simulation box, demonstrating that the method used to compute the profiles is not biased.

We have similarly computed the average galaxy density radial profile of filaments detected in the Magneticum, TNG300-2, TNG100-2 and Illustris-2 simulations. The results are shown in the lower panel of Fig. \ref{Fig:profile3_nulltest}, respectively in red, blue, green and yellow, along with the reference profile that is shown in black. For the sake of comparison with the profile obtained from the TNG300-1 simulation, we have rescaled all the profiles by their respective background densities.

We can see that the TNG300-1, Magneticum, TNG300-2 and TNG100-2 profiles are essentially the same, despite some differences that will be discussed in the next paragraph. The Illustris-2 profile is significantly different, particularly near the core of the filament. The shallow core of Illustris-2 might be caused by AGN and possible other feedback effects that are known to be more important in this simulation, and that have already been identified as responsible of the excessively high depletion rate of gas in massive haloes at low redshift, which is in disagreement with observations \citep{Genel2014ILLUSTRIS, Haider2016ILLUSTRIS}. To confirm this hypothesis, we computed galaxy density profiles in nodes in the Illustris-2 and TNG100-2 simulation (assuming spherical symmetry). We recall that nodes are identified by the maximum density critical points, and these objects are masked in the analysis of filaments (see Sect.~\ref{SubSect_radial_density_profiles_METHOD}). The resulting density profiles in nodes are shown in Fig.~\ref{Fig:HALOprofiles} of Appendix~\ref{Appendix:AGN}, and they present the same trend as the profiles of filaments, i.e. the Illustris-2 slope is shallower than the TNG100-2 one. These results support the fact that the densest regions of the cosmic web, like massive haloes and cores of filaments, are particularly sensitive to feedback effects.
However, it is important to note that, while baryonic physics plays a role in the distribution of matter around filaments, the essential and dominant driver remains gravity (as expected), as is shown by the strong similarity of the density profiles of the other simulations (cf. Fig.~\ref{Fig:profile3_nulltest}), even though their feedback models (TNG and Magneticum models) are not identical.

Despite the good visual agreement in the other profiles of Fig.~\ref{Fig:profile3_nulltest}, let us now discuss the slight deviations spotted at radial distances of $\sim 1$ Mpc. Indeed, we note a softer decrease of galaxy density from the core to the outskirts of the filament in the TNG300-2 profile (blue) with respect to the TNG300-1 reference profile (black), yielding a more extended and fatter core in the low resolution run (see Table~\ref{Table:sims_overview}).
This effect could be explained by a reduced precision in the detection of filaments by the DisPerSE algorithm, coming from the smaller number of tracers (i.e. galaxies) in lower resolution simulations. In order to verify this hypothesis, we thin the TNG300-1 galaxy catalogue by randomly removing every other galaxy so that the density of tracers matches that of TNG300-2 (see Table~\ref{Table:sims_overview}). We then extract the DisPerSE skeleton of this thinned catalogue using the same DisPerSE parameters as presented in Sect.~\ref{SubSec:Disperse}, and we compute the galaxy density profiles of the thinned catalogue around these newly extracted filaments. The resulting profile is displayed in Fig.~\ref{Fig:resolution} (green dashed curve with triangles) and shows a perfect agreement with the TNG300-2 profile (in blue). We specify that the left panel of this figure displays the rescaled density profiles, while the right panel shows the profiles at their respective galaxy backgrounds (denoted by the thin horizontal lines). In addition, we also analyse the distribution of galaxies of the thinned TNG300-1 catalogue around the original TNG300-1 reference filaments. This results in the pink dashed profile with squares, which has exactly the same shape as the TNG300-1 reference (in black). We note that this curve perfectly overlaps with the reference in the left panel of rescaled profiles, proving that the distribution of matter around filaments is not changed with resolution. Hence, these tests show that the fatter core of the TNG300-2 profile is purely due to DisPerSE precision effects that increase the uncertainty in the position of filaments.\\

The Magneticum profile (red curve) of Fig.~\ref{Fig:profile3_nulltest} also shows these resolution effects. Indeed, the Magneticum simulation has a resolution similar to TNG300-2 (see Table~\ref{Table:sims_overview}), so their density profiles are in good agreement. Still, we note that this simulation has also a different cosmology and different physical models than the Illustris-TNG series, but here we are not able to disentangle these effects in the density profiles. However, despite the differences in simulations, the profiles of TNG300-1 and Magneticum are remarkably similar from a statistical point of view, as confirmed by the $p$-value of 0.22 obtained from the two-sample Kolomogorov-Smirnov test.


\subsection{\label{SubSect:bybinsL}Filament populations}

\begin{figure}
   \centering
   \includegraphics[width=0.5\textwidth]{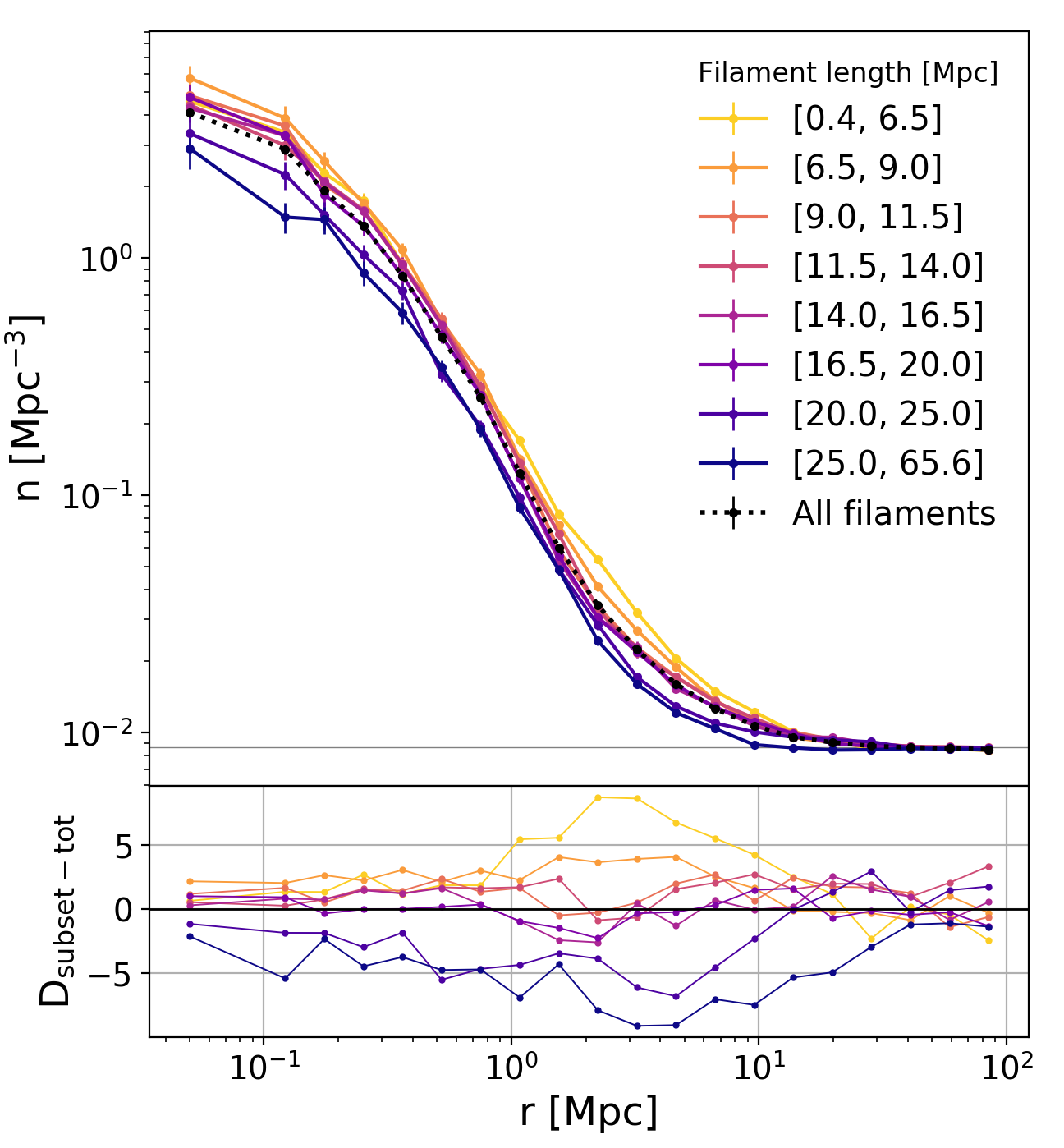}
   \caption{\textit{Top:} Radial galaxy density profiles of filaments by bins of filament length for the reference catalogue (TNG300-1). The black curve corresponds to the average of all the filaments, regardless of their length, as presented in Sect.~\ref{Subsect:radial_density_profile}. The coloured curves correspond to the average of length-selected filaments. The gray horizontal line represents the background galaxy density of the simulation box. 
   \textit{Bottom:} Deviation from the mean $D_{\mathrm{subset} - \mathrm{tot}}$ of the length-selected profiles (see definition in Eq.~\ref{Eq:delta}).}
    \label{Fig:radial_profiles_byL}
    \end{figure}

The average galaxy density profiles obtained in Sect. \ref{Subsect:radial_density_profile} characterise the entire population of filaments. In order to explore possible dependencies with the filament lengths, we split the  TNG300-1 filament catalogue in eight different bins of length chosen so that each contains the same number of segments, $\sim 2200$. We recall that the profile of filaments in a given bin of length is the average of the profiles of the segments forming the filaments in this same bin. We show, in the top panel of Fig.~\ref{Fig:radial_profiles_byL}, the average profile of all the filaments (black dotted curve) together with the galaxy density profiles in the eight bins of length (from yellow to blue). We can see that filaments of different lengths have significantly different radial galaxy densities. The profiles of the shortest filaments lie above the average at all radial distances, whereas the longest filaments are found below the average.
The bottom panel shows the deviation from the average profile for each bin of length, which we define as 
    \begin{equation}
        D_{\mathrm{subset} - \mathrm{tot}} = \frac{n_{\mathrm{subset}} - n_\mathrm{tot}}{\sqrt{\sigma^2_\mathrm{subset} + \sigma^2_\mathrm{tot}}},
        \label{Eq:delta}
    \end{equation}
where $n_\mathrm{subset}$ is the profile of the filaments in a given bin, $n_\mathrm{tot}$ is the average profile (as in Sect.~\ref{Subsect:radial_density_profile}), $\sigma_\mathrm{subset}$ and $\sigma_\mathrm{tot}$ are the bootstrap errors of the corresponding profiles.

While the filaments of lengths in the range $9 \leq L_\mathrm{f} < 20$ Mpc do not present significant deviations from the average profile ($|D_{\mathrm{subset} - \mathrm{tot}}| < 2$), the shortest and longest bins of length deviate by more than $2\sigma$, and the differences are the most significant ($|D_{\mathrm{subset} - \mathrm{tot}}| \sim 8$) for distances between 1 to 10 Mpc from the filament spine. 
We therefore split our catalogue into two populations: short filaments with lengths shorter than 9 Mpc, and long filaments with lengths longer than 20 Mpc.\\

We compute the average profiles of all the short ($L_\mathrm{f} < 9$ Mpc) and long filaments ($L_\mathrm{f} \geq 20$ Mpc), respectively $\cal{S}$ and $\cal{L}$ profiles hereafter.
Despite a larger number of short filaments (2846 vs 611), the average $\cal{S}$ and $\cal{L}$ profiles are computed from approximately the same number of segments (respectively 4733 and 4129). We therefore note that short filaments are statistically made of less segments than long filaments: the mean number of segments per filament are $\bar{N}_\mathrm{seg}^{\cal{S}} = 1.7$ for short and $\bar{N}_\mathrm{seg}^{\cal{L}} = 6.5$ for long.

In the following, we study short and long filaments by using Kolmogorov-Smirnov (KS) statistical tests. 
First, we compare the measured $\cal{S}$ (resp. $\cal{L}$) profile with an average bootstrap profile computed from the complementary set of segments, i.e. the subset of profiles whose segments do not belong to the $\cal{S}$ (resp. the $\cal{L}$) population. We use the KS to test the null hypothesis which assumes that the $\cal{S}$ and the related bootstrap profiles (resp. the $\cal{L}$ and the related bootstrap profiles) are drawn from the same parent distribution. We repeat this procedure 1000 times in order to have a distribution of $p$-values that takes into account the statistical properties of the complementary population. The resulting  $p$-values are found to be always smaller than $6 \times 10^{-3}$ and $4 \times 10^{-6}$ for the short and long populations, respectively. These results indicate that the density profiles of short and long filaments can not be randomly reproduced, and therefore strongly suggest that there are at least two populations in the total set of filaments.
Finally, a unique two-sample KS test is performed to directly compare the $\cal{S}$ and $\cal{L}$ profiles. The resulting $p$-value of $5.9 \times 10^{-11}$ confirms that these filaments indeed constitute two different populations. \\

We compute the average short and long profiles of the Magneticum, TNG300-2 and TNG100-2 filaments following the same procedure, i.e. for each simulation we split the filament catalogue in thin bins of length having the same number of segments, and we define the $\cal{S}$ and $\cal{L}$ populations from the subset profiles that deviate by more than $2 \sigma$ from the total average profile.
We note that the significance of the deviations of Eq.~\ref{Eq:delta} strongly depends on the statistics of the catalogue, as the errors on the average profiles are sensitive to the number of filaments in each simulation ($\sigma \propto \sqrt{N}$). In order to take into account these effects and to be able to compare the deviations of catalogues with different statistics, we rescaled the errors of the simulations to those of TNG300-1 by replacing $\sigma$ with $\sigma \times \sqrt{N/N^\mathrm{ref}}$ in Eq.~\ref{Eq:delta}. Here $N$ denotes the number of segments either in the subset bin or in the total catalogue of the corresponding simulation, and $N^\mathrm{ref}$ corresponds to the same quantities in TNG300-1, the reference simulation. 
The resulting lengths defining the $\cal{S}$ and $\cal{L}$ populations are shown in Table~\ref{Table:SLbins}. As expected, the length boundaries for the TNG300-2 profiles ($14.4$ and $27.2$ Mpc) differ the most from the reference values ($9$ and $20$ Mpc), and this is due to the significantly different filament length distributions of these simulations (as discussed in Sect.~\ref{SubSec:filament_catalogue}.)\\

Concerning the Illustris-2 filaments, we found the surprising result that their galaxy densities showed no dependency whatsoever with filament length. Indeed, the splitting of this catalogue in thin bins of length resulted in profiles that closely followed the total average (presented in Fig.~\ref{Fig:profile3_nulltest} and discussed in Sect.~\ref{Subsect:radial_density_profile}). The shortest and longest bins did not show the expected trend (clearly exhibited by all the other simulations) and all the corresponding deviations displayed an oscillation around zero. This result is not due to the lower statistics of the Illustris-2 box, as the populations of filaments were clearly detected in the TNG100-2 simulation (see text above and Table~\ref{Table:SLbins}), whose box contains almost the same number of filaments as Illustris-2. Most likely, this surprising founding might be related with the specific model of baryonic physics of this simulation \citep{Genel2014ILLUSTRIS, Haider2016ILLUSTRIS}, which we discussed in Sect.~\ref{Subsect:radial_density_profile}. This may be a hint that different populations of filaments are not only a natural result of cosmological accretion, but they also rely on the physics of baryons, which shapes the distribution of matter around filaments to a certain extent.

\begin{figure}
    \centering
    \includegraphics[width=0.5\textwidth]{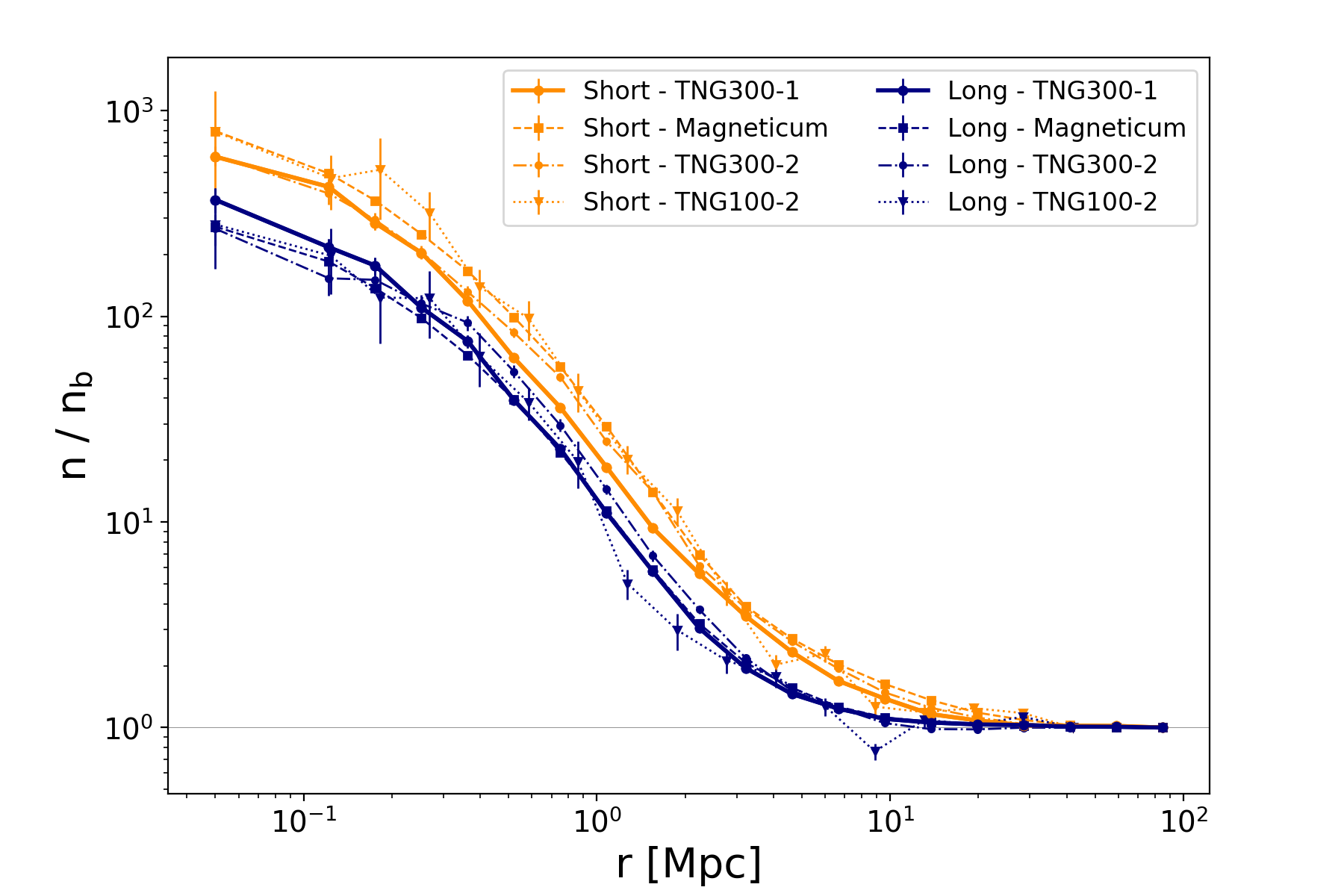}
    \caption{Average galaxy density profiles around short (orange curves) and long filaments (in blue) for the reference TNG300-1 simulation (thick solid lines with dots), and the Magneticum (dashed lines with squares), TNG300-2 (dotted-dashed lines with points) and TNG100-2 (dotted lines with triangles) simulations. The length limits defining short and long filaments in each simulation are presented in Table~\ref{Table:SLbins}. For the sake of comparison, all the profiles have been rescaled by their respective background.}
    \label{Fig:pro_SL_allsims}%
    \end{figure}

That being said, Fig.~\ref{Fig:pro_SL_allsims} presents the derived $\cal{S}$ and $\cal{L}$ profiles of filaments in the TNG300-1, TNG300-2, Magneticum and TNG100-2 simulations. Orange and blue colours denote respectively short and long filaments. We see that the $\cal{S}$ and $\cal{L}$ profiles of these four simulations are remarkably compatible, modulo some resolution effects in the TNG300-2 and Magneticum filaments, causing the broadening of the cores (as discussed in Sect.~\ref{Subsect:radial_density_profile}). 
In addition, the Magneticum profile also shows slightly higher density values at the core of the short filaments, which might be due to the different (but compatible) physical models of Magneticum \citep{SpringelDiMatteo2005_feedbackmodels, Fabjan2010_AGNfeedback} with respect to the TNG series \citep{Pillepich2018TNGmodel}. 
Finally, the small deviations of the TNG100-2 filaments with respect to the other profiles may be caused by the very reduced statistics in this simulation box. Indeed, the short and long profiles in TNG100-2 result from the average of only $\sim 200$ segments (vs 2200 in the reference TNG300-1), so the mean values may not be representative of the underlying density distribution.

\begin{table}
\caption{Definition of short and long filaments for the different simulations analysed in this work.}
\label{Table:SLbins}     
\centering  
\begin{tabular}{ c  c  c }
 \hline\hline    
   Simulation & Short [Mpc] & Long [Mpc] \\ \hline
   \textbf{TNG300-1} &  $L_\mathrm{f} < 9.0$ & $ L_\mathrm{f} \geq  20.0$  \\
   Magneticum & $L_\mathrm{f} < 10.7$ & $L_\mathrm{f} \geq 16.5$ \\
   TNG300-2 & $L_\mathrm{f} < 14.4$ & $L_\mathrm{f} \geq 27.2$ \\
   TNG100-2 & $L_\mathrm{f} < 8.0$ & $L_\mathrm{f} \geq 18.8$ \\
  \hline
 \end{tabular}
\end{table}

We compute some characteristic radial scales for the TNG300-1 profiles. First, the radial extent of the $\cal{L}$ profile, $r_\mathrm{e}^{\cal{L}} \sim 19$ Mpc, computed with respect to the corresponding null-test, is found to be almost twice as small as that of the $\cal{S}$ profile, $r_\mathrm{e}^{\cal{S}} \sim 35$ Mpc. Likewise, the radius $r_2$, defined as the radius for which the galaxy density is twice the value of the  background density $n_\mathrm{b}$, is smaller for long ($r_{2}^{\cal{L}} \sim 3$ Mpc) than for short filaments ($r_{2}^{\cal{S}} \sim 5$ Mpc).
Qualitatively, we find similar results in the other simulations, namely that long filaments are radially less extended than short filaments, and thinner at all radial scales.

\subsection{\label{SubSect:connetionCPmax}Environments of short and long filaments}

With the aim of probing the environment of these two different populations, we compute the galaxy density excess. This quantity, hereafter noted $\langle 1+\delta \rangle_{r<R}$, corresponds to the excess of density of a filament with respect to the background density, summed up to a radial distance of $R$, in Mpc. It is defined as:
 \begin{equation}
     \langle 1+\delta \rangle_{r<R} \, \, =  \frac{1}{R} \, \int_{0}^{R} \frac{n_\mathrm{fil}(r)}{n_\mathrm{b}} \, \mathrm{d}r.
     \label{Eq:overdens}
 \end{equation}
In this equation, $n_\mathrm{fil}(r)$ is the filament galaxy density profile, which corresponds to the average profile of its segments, and $n_\mathrm{b}$ corresponds to the background density of the simulation.

\begin{figure}
   \centering
   \includegraphics[width=0.5\textwidth]{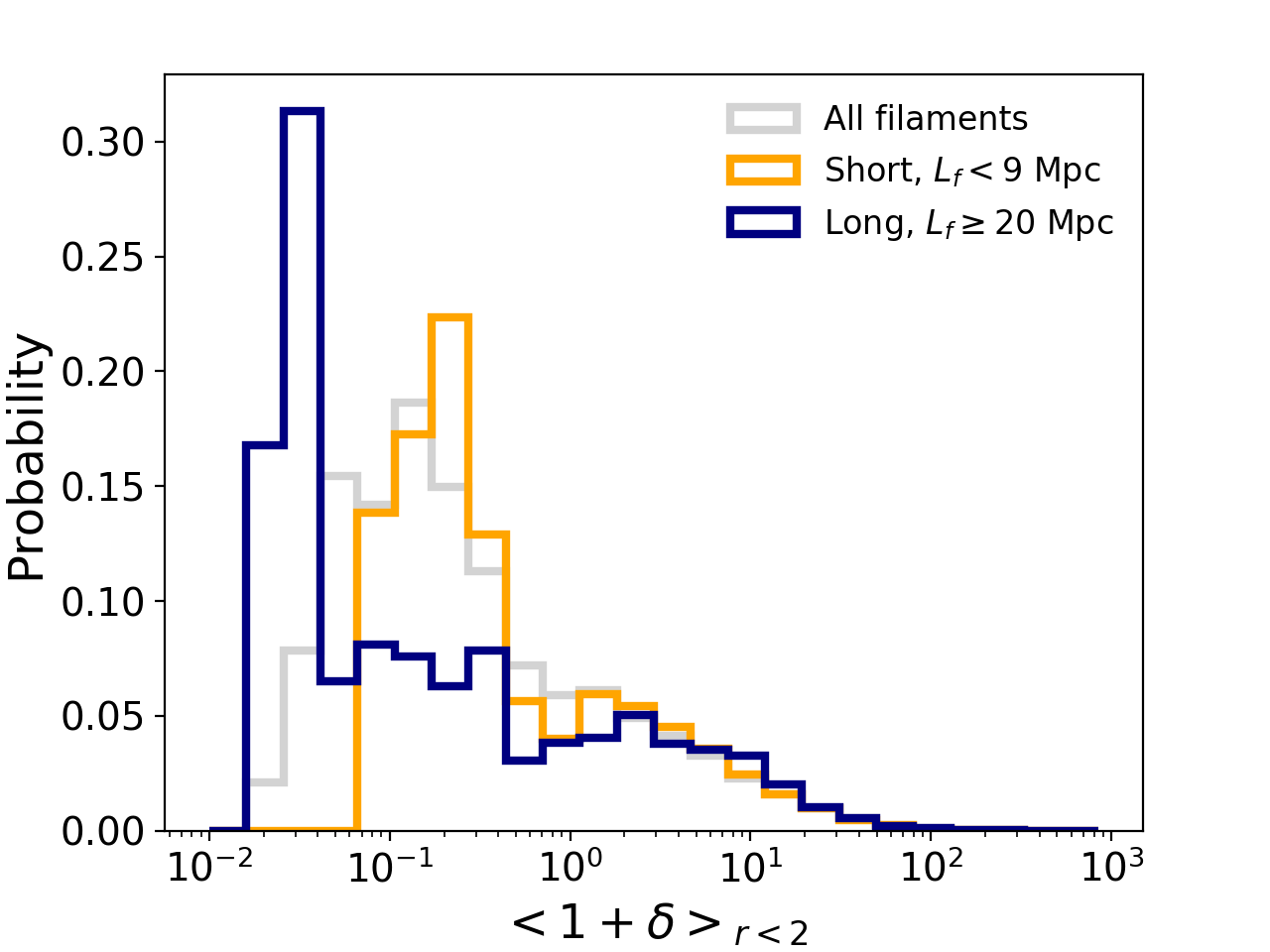}
   \caption{Probability distributions of the excess of density $\langle 1+\delta \rangle_{r<2}$ (see Eq.~\ref{Eq:overdens}) of short (orange), long (blue) and the total population of filaments (gray). This quantity corresponds to the excess of density of a filament with respect to the  background density, averaged up to a radial distance of $R=2$ Mpc.}
    \label{Fig:overdensities}
    \end{figure}

We compute this quantity for filaments of the reference catalogue TNG300-1.
The resulting $\langle 1+\delta \rangle_{r<2}$ probability  distributions of the short and long filaments are shown in Fig.~\ref{Fig:overdensities}. The radial scale of integration, $R=2$ Mpc, was chosen to correspond to the radius of filaments found in other studies (see for example \citealp{AragonCalvo2010, Colberg2005, GonzalezPadilla2010, Bond2010b, Cautin2014}), although we note that other analyses have found different filament radii \citep[e.g.][]{Bonjean2019filaments, Tanimura2020_byopic}. The distribution for the total population of filaments is also displayed for comparison. It is important to note that the area under each histogram is normalised to one by dividing the counts by the number of measurements times the bin width.

First, we observe that the density excess of filaments spans almost 5 orders of magnitude, from very under-dense filaments $\langle 1+\delta \rangle_{r<2} \, \sim 1.9 \times \, 10^{-2}$, to filaments with densities of the order of clusters $\langle 1+\delta \rangle_{r<2} \, \sim 8.5 \times \, 10^2$.
These results are in good agreement with those of \cite{Cautin2014} where the densities of filaments, detected in a DM simulation, are also found to span a large range, similar to our findings.

Let us now focus on the specific features of the distributions of the $\cal{S}$ and $\cal{L}$ populations. We can see that these distributions share a common regime at $\langle 1+\delta \rangle_{r<2} \sim [2-300]$, but they are significantly different at lower values. Indeed, the distribution of long filaments exhibits a clear excess towards the lowest values (peak at $\langle 1+\delta \rangle_{r<2} \sim 0.03$) whereas the distribution of short filaments is shifted towards higher values (peak at $\langle 1+\delta \rangle_{r<2} \sim 0.2$) and its minimum is 0.07, which is more than twice the value for long filaments, 0.02. Similarly, towards the highest values, the distribution for the short filaments reaches a maximum density of 848, which is a factor of two larger than the maximum of long ones, 415. The mean values are 41 and 34, respectively for the $\cal{S}$ and $\cal{L}$ populations, and the medians are 13 and 14. These results indicate that the density distributions of short and long filaments are different, with the statistical trend that short filaments are on average denser than long ones.

\begin{figure}
   \centering
   \includegraphics[width=0.5\textwidth]{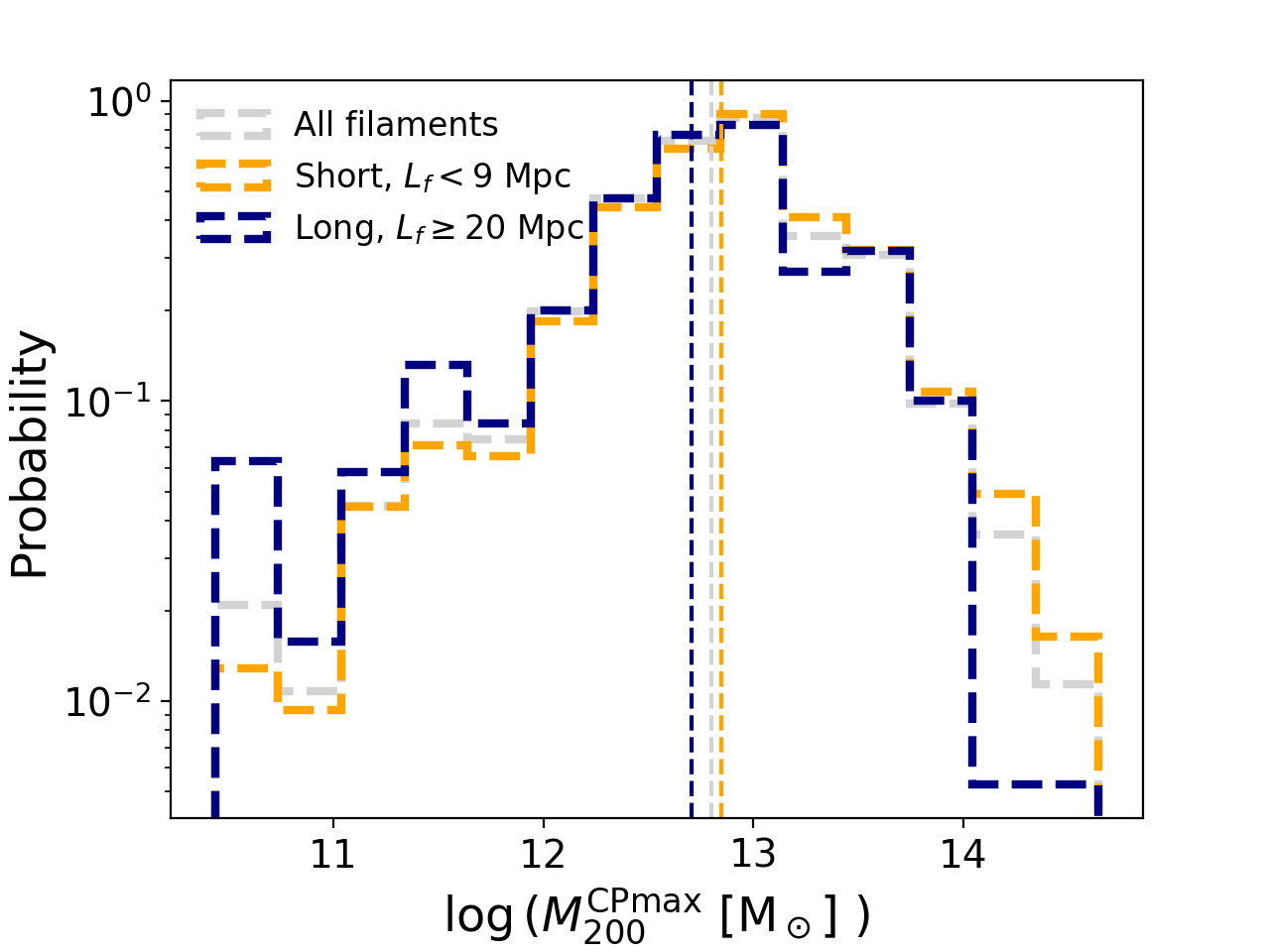}
   \caption{Distribution of DM masses $M^{\mathrm{CPmax}}_{200}$ of the maximum density critical points (CPmax) in the TNG300-1 simulation. CPmax connected to short (resp. long) filaments are presented in orange (resp. blue). The vertical dashed lines correspond to the mean values of the distributions.}
    \label{Fig:mass_node_SL}
    \end{figure}

In addition to the galaxy density excess, we analyse the connection of short and long filaments to their maximum density critical point (CPmax) in the reference, TNG300-1, simulation. 
The CPmax are found using the DisPerSE code (see Sect.~\ref{SubSec:Disperse}) and they correspond to the points of maximum density defining one of the two extremities of a filament. In the previous analyses, they have been masked to suppress the contribution of the nodes of the cosmic web in the radial density profiles (Sect.~\ref{SubSec:filament_catalogue}). 
We compute the masses at the CPmax positions, hereafter called $M^{\mathrm{CPmax}}_{200}$, defined as the DM mass enclosed in a sphere of radius $R_{200}$. The resulting mass distribution is displayed in Fig.~\ref{Fig:mass_node_SL} for short, long and all the filaments (in orange, blue, and gray, respectively).

We find that short filaments may be connected to more massive objects (maximum mass $\sim 10^{14.65} \, \mathrm{M}_\odot$) than long filaments (max $\sim 10^{14.05} \, \mathrm{M}_\odot$).
The mean values of these distributions also reflect this trend, the values are $10^{12.71}$ and $10^{12.85} \mathrm{M}_\odot$ for the long and short populations, respectively.

These studies, combined with the density profiles of the previous section, all suggest that short and long filaments trace different environments of the cosmic web. Short filaments are puffier, denser, radially more extended than long filaments and possibly connected to more massive structures. They may thus be embedded in over-dense environments at the proximity of clusters and might correspond to bridges of matter between over-dense structures. On the contrary, long filaments are thinner, less dense and connected on average to less massive objects. This long and thinner population may represent the cosmic filaments shaping the structure of the cosmic web and lying in under-dense regions (e.g. near cosmic voids).

\subsection{\label{Sect:Fits} Models for the galaxy density around filaments}

    \begin{figure}
    \centering
    \includegraphics[width=0.5\textwidth]{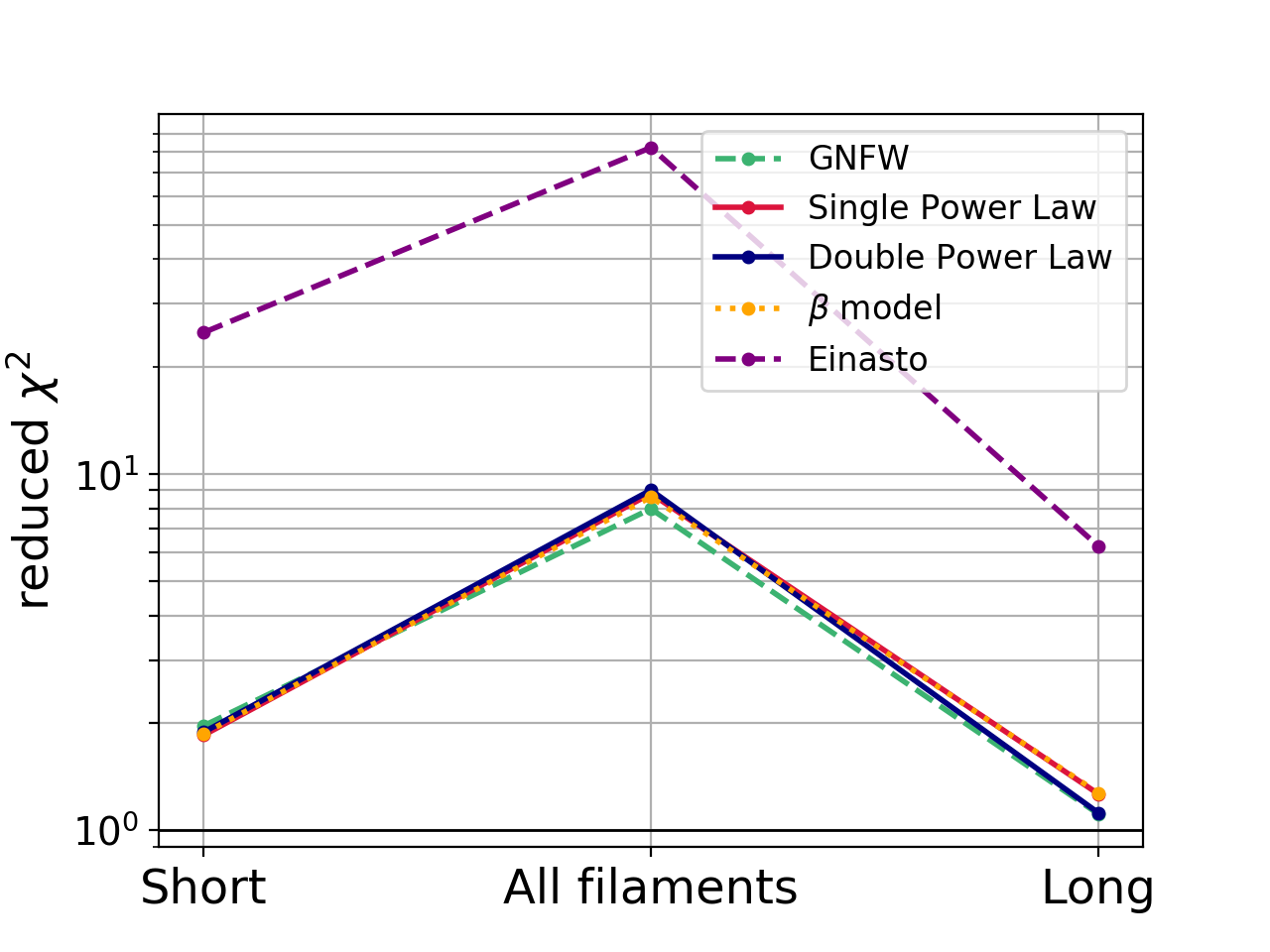}
    \caption{$\chi^2_\nu$ results of the fitting of different models to short, all and long filaments. Each colour represents a different model. We can clearly see a segregation between the short and long populations: when mixed together, they are less well fitted than when taken separately.}
    \label{Fig:chisquare}%
    \end{figure}
    
Several analytic models can be proposed for the radial profiles of galaxy densities. Some of these  are borrowed from cluster literature, like the generalised Navarro, Frenk and White model \citep[GNFW,][]{Hernquist1990,Navarro1997_NFW, Nagai2007,Arnaud2010}, presented in Eq.~\ref{Eq:GNFW} , 
    \begin{equation}\label{Eq:GNFW}
    n^{\mathrm{GNFW}}(r) = \frac{n_0}{\left( \frac{r}{r_\mathrm{f}} \right)^\alpha \, \left[ 1+ \left( \frac{r}{r_\mathrm{f}} \right)^\gamma \right]^{\frac{\beta - \alpha}{\gamma}} } + n_\mathrm{b},
    \end{equation}
    the Einasto model \citep{Einasto1965, Ludlow2017einasto} of Eq. \ref{Eq:Einasto},
    \begin{equation}\label{Eq:Einasto}
    n^{\mathrm{Einasto}}(r) = n_0 \exp \left[-\left(\frac{r}{r_\mathrm{f}} \right)^a \right] + n_\mathrm{b}
    \end{equation}
     and $\beta$ models of Eq. \ref{Eq:Beta} \citep{Cavaliere1976, Arnaud2009, Ettori2013}.
    \begin{equation}\label{Eq:Beta}
    n^{\mathrm{\beta}}(r) = \frac{n_0}{\left[ 1+ \left( \frac{r}{r_\mathrm{f}} \right)^\alpha \right]^\beta} + n_\mathrm{b}.
    \end{equation}

Other models are more general and empirical expressions proposed to describe filaments, like the double power law \citep{AragonCalvo2010} of Eq.~\ref{Eq:2PL}, hereafter called PL2, 
    \begin{equation}\label{Eq:2PL}
    n^{\mathrm{PL2}}(r) = \frac{n_0}{ \left( \frac{r}{r_\mathrm{f}} \right)^\alpha + \left( \frac{r}{r_\mathrm{f}} \right)^\beta} + n_\mathrm{b}
    \end{equation}
or simply a single power law (PL1) of Eq.~\ref{Eq:1PL} \citep{Colberg2005},
    \begin{equation}\label{Eq:1PL}
    n^{\mathrm{PL1}}(r) = \frac{n_0}{ 1+ \left( \frac{r}{r_\mathrm{f}} \right)^\beta} + n_\mathrm{b}.
    \end{equation}

These models usually describe two regimes: small ($r \ll r_\mathrm{f}$) and large ($r \gg r_\mathrm{f}$) radii (where $r_\mathrm{f}$ is a characteristic radius), with a possibility of a transitional region between them in the GNFW case determined by the parameter $\gamma$. Each model is characterised by slopes describing the small and large radii regimes. Notice that the names of the exponents in the formulae above  have been chosen with the aim of easing the comparison of the models, and may not correspond to the usual labelling.

\bigskip
We fitted the average profile of the galaxy density, obtained for the total filament population, for the short and long filaments of the TNG300-1 simulation, to these analytic models and we display in Fig. \ref{Fig:chisquare} the reduced chi-squared values, $\chi^2_\nu$. We first note that the fit to the Einasto profile (purple dashed line) always yields the highest $\chi^2_\nu$ values indicating that it is not an adequate model to describe the galaxy density profiles.
We also observe that the fits perform always worse ($\chi^2_\nu\sim 9$) for the entire filament set than for the short or the long filament populations considered separately (with respectively $\chi^2_\nu\sim 1.9$ and $\sim 1.2$), as expected.
We point out that all the tested models perform equally well, in terms of their $\chi^2_\nu$ values in Fig.~\ref{Fig:chisquare}, for the short or the long filaments populations.

\begin{table*} 
 \caption{Results of the MCMC fits on short filaments ($L_{f}<9$ Mpc). The $\cal{S}$ galaxy density profile is adjusted using the models described in Eq.~\ref{Eq:GNFW}, ~\ref{Eq:Beta}, \ref{Eq:2PL} and \ref{Eq:1PL}.}
 \label{Table:fitSHORT}    
\centering          
 \begin{tabular}{ c  c  c  c  c  c  c }
 \hline\hline  
     & $n_0$ [$\mathrm{Mpc}^{-3}$] & $r_\mathrm{f}$ [Mpc] & $\alpha$ & $\beta$ & $\gamma$ & $\chi^2_\nu$ \\ 
    \hline
    GNFW & $2.80_{-0.726}^{+1.36}$ & $0.21_{-0.04}^{+0.04}$ & $0.41_{-0.25}^{+0.18}$ & $1.80_{-0.02}^{+0.02}$ & $3.54_{-1.03}^{+2.64}$ & \textbf{1.96} \\
    $\beta$-profile & $5.36_{-0.65}^{0.75}$ & $0.15_{-0.01}^{+0.01}$ & $2.18_{-0.28}^{+0.38}$ & $0.84_{-0.13}^{+0.13}$ & - & \textbf{1.87} \\
    PL2 & $6.69_{-0.71}^{+0.76}$ & $0.13_{-0.01}^{+0.01}$ & $-0.18_{-0.16}^{+0.15}$ & $1.82_{-0.02}^{+0.02}$ & - &\textbf{1.88} \\
    PL1 & $6.07_{-0.48}^{+0.52}$ & $0.14_{-0.01}^{+0.01}$ & - &  $1.82_{-0.01}^{+0.01}$ & - & \textbf{1.85} \\
  \hline
 \end{tabular}
\end{table*}

\begin{table*} 
 \caption{Results of the MCMC fits on long filaments ($L_{f} \geq 20$ Mpc). The $\cal{L}$ galaxy density profile is adjusted using the models described in Eq.~\ref{Eq:GNFW}, ~\ref{Eq:Beta}, \ref{Eq:2PL} and \ref{Eq:1PL}.}
 \label{Table:fitLONG}    
\centering          
 \begin{tabular}{ c  c  c  c  c  c  c }
 \hline\hline  
     & $n_0$ [$\mathrm{Mpc}^{-3}$] & $r_\mathrm{f}$ [Mpc] & $\alpha$ & $\beta$ & $\gamma$ & $\chi^2_\nu$ \\ 
    \hline
    GNFW & $1.15_{-0.35}^{+0.91}$ & $0.31_{-0.06}^{+0.06}$ & $0.54_{-0.31}^{+0.19}$ & $2.10_{-0.04}^{+0.06}$ & $2.63_{-0.79}^{+1.42}$ & \textbf{1.11} \\
    $\beta$-profile & $3.15_{-0.54}^{+0.72}$ & $0.21_{-0.02}^{+0.02}$ & $1.60_{-0.25}^{+0.31}$ & $1.35_{-0.24}^{+0.29}$ & - & \textbf{1.27} \\
    PL2 & $1.96_{-0.35}^{+0.37}$ & $0.25_{-0.03}^{+0.04}$ & $0.28_{-0.14}^{+0.13}$ & $2.14_{-0.04}^{+0.04}$ & - & \textbf{1.12} \\
    PL1 & $2.50_{-0.24}^{+0.26}$ & $0.21_{-0.01}^{+0.02}$ & - & $2.09_{-0.03}^{+0.03}$ & - & \textbf{1.26} \\
  \hline
 \end{tabular}
\end{table*}

\begin{table*} 
 \caption{Results of the MCMC fits on the total population of filaments. The total galaxy density profile is adjusted using the models described in Eq.~\ref{Eq:GNFW}, ~\ref{Eq:Beta}, \ref{Eq:2PL} and \ref{Eq:1PL}.}
 \label{Table:fitALL}    
\centering          
 \begin{tabular}{ c  c  c  c  c  c  c }
 \hline\hline  
     & $n_0$ [$\mathrm{Mpc}^{-3}$] & $r_\mathrm{f}$ [Mpc] & $\alpha$ & $\beta$ & $\gamma$ & $\chi^2_\nu$ \\ 
    \hline
    GNFW & $1.67_{-0.19}^{+0.26}$ & $0.25_{-0.02}^{+0.02}$ & $0.57_{-0.08}^{+0.07}$  & $1.87_{-0.01}^{+0.01}$ & $4.94_{-0.98}^{+1.59}$  & \textbf{7.97} \\
    $\beta$-profile & $3.93_{-0.25}^{+0.27}$ & $0.16_{-0.01}^{+0.01}$ & $2.43_{-0.17}^{+0.19}$  & $0.77_{-0.06}^{+0.06}$ & - & \textbf{8.60} \\
    PL2 & $5.18_{-0.29}^{+0.30}$ & $0.14_{-0.01}^{+0.01}$ & $-0.17_{-0.08}^{+0.07}$  & $1.89_{-0.01}^{+0.01}$ & - &\textbf{9.00} \\
    PL1 & $4.69_{-0.20}^{+0.20}$ & $0.16_{-0.01}^{+0.01}$ & - & $1.90_{-0.01}^{+0.01}$ & - & \textbf{8.80} \\
  \hline
 \end{tabular}
\end{table*}

In the following, we focus on the parameters of the fits (fixing $n_\mathrm{b}$ to the background galaxy density of the simulation box) that we explore using Monte Carlo Markov Chains (MCMC) for the short, long and total populations. We discard the Einsato model and present the best-fit parameters (namely the median values of the posterior distributions) in Tables \ref{Table:fitSHORT} and \ref{Table:fitLONG} for the short and long filaments respectively. The parameters fitting the entire population are presented in Table \ref{Table:fitALL}.

\begin{figure*}
    \centering
    \includegraphics[width=0.5\textwidth]{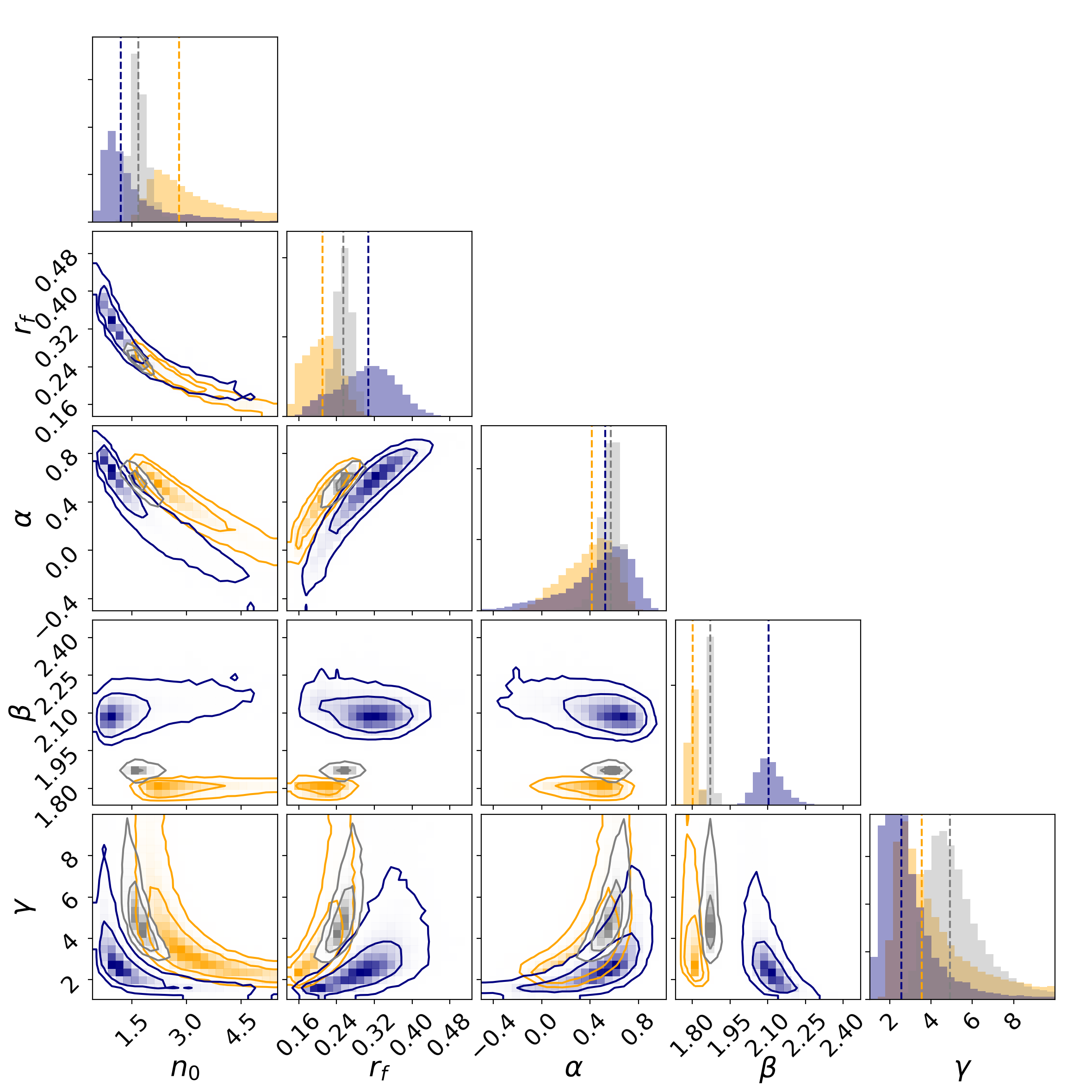}\includegraphics[width=0.5\textwidth]{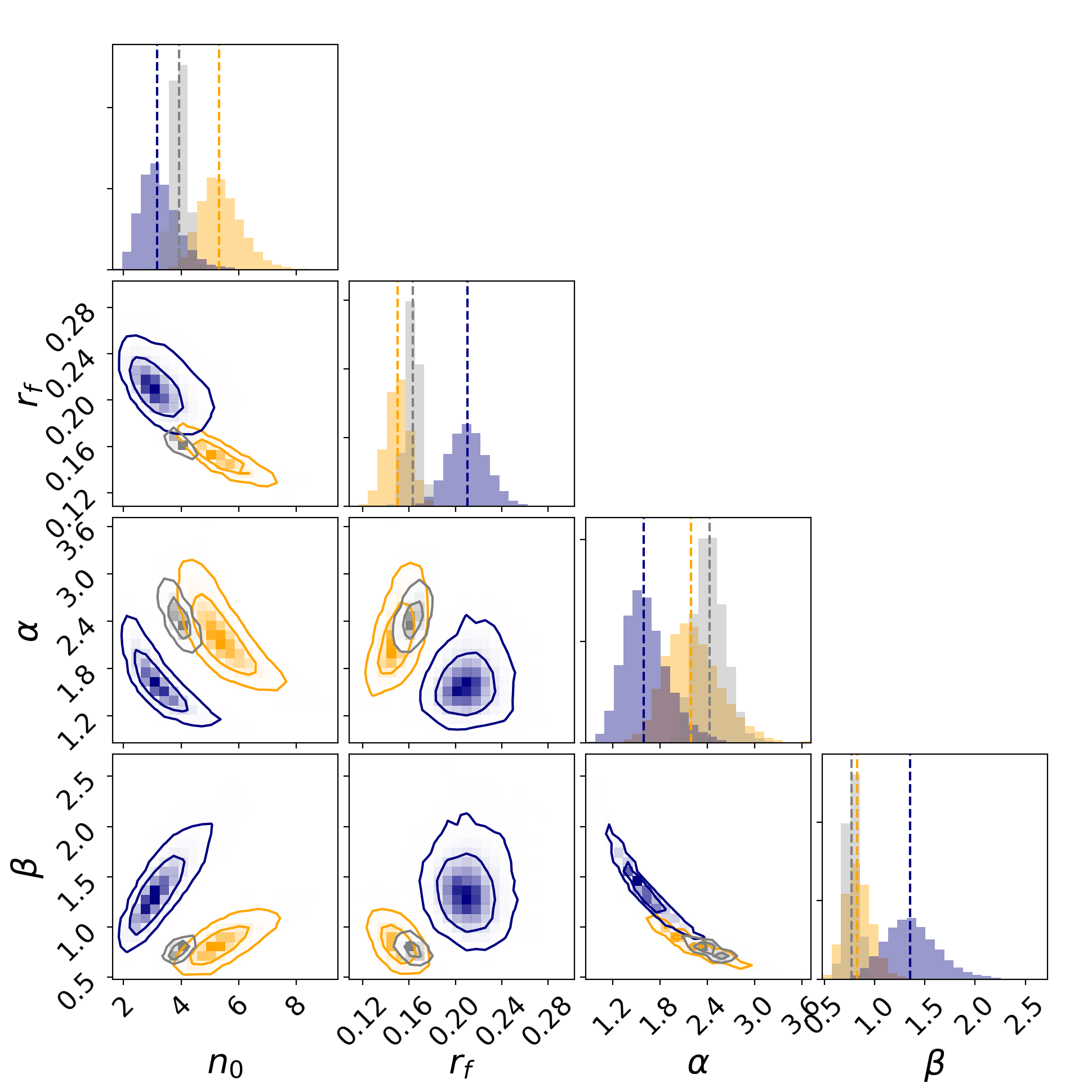}
    \includegraphics[width=0.5\textwidth]{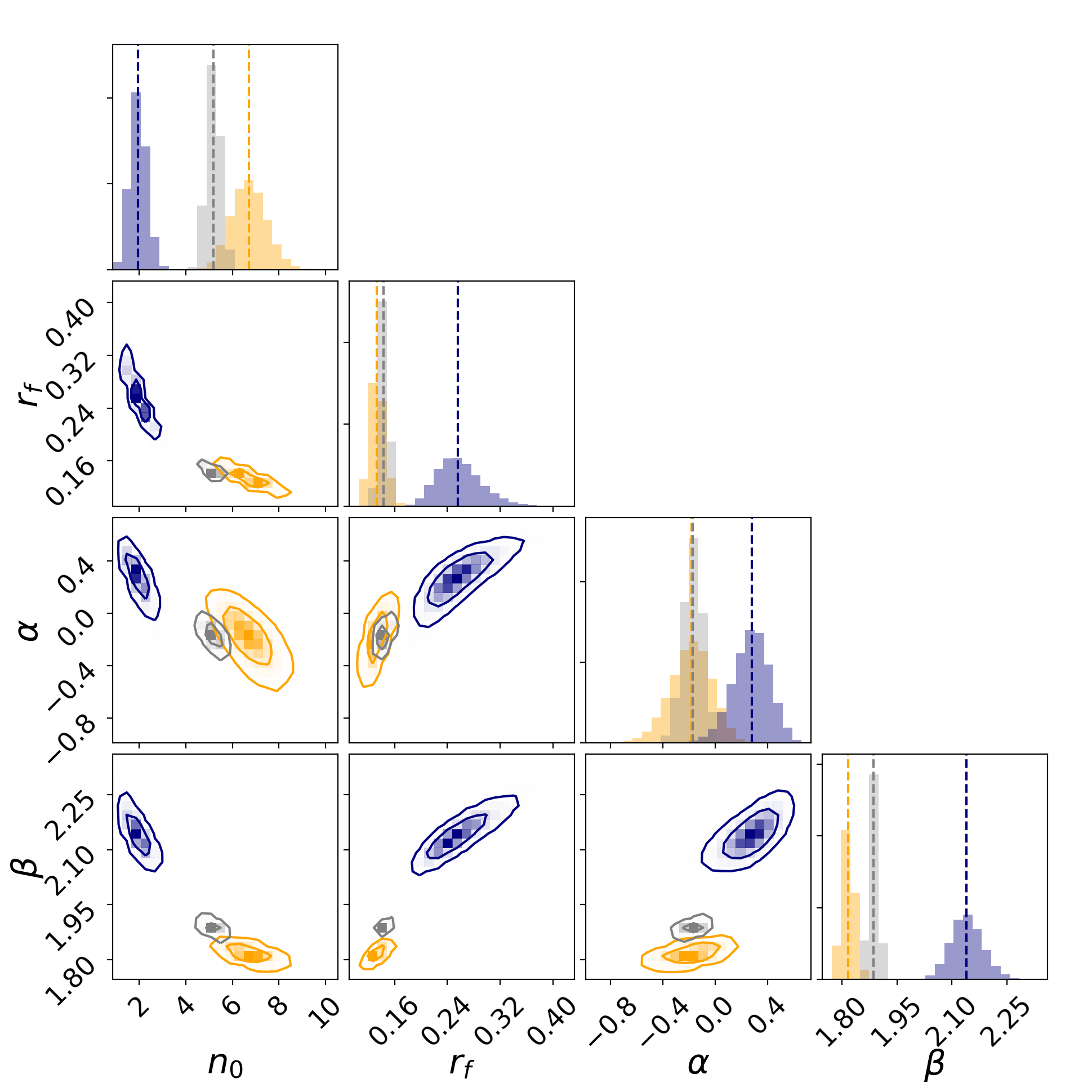}\includegraphics[width=0.5\textwidth]{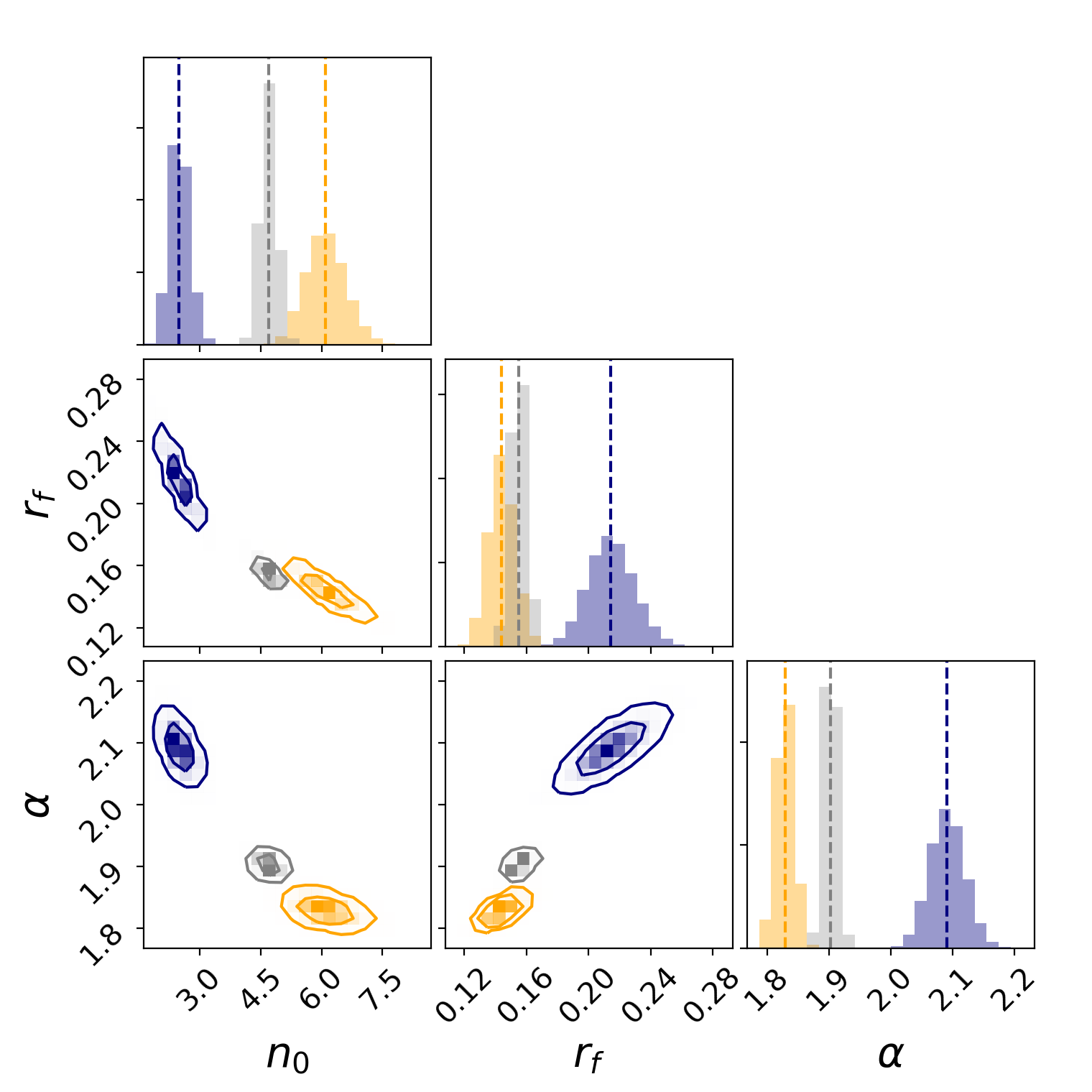}
    \caption{MCMC exploration of the distribution of parameters of four different models fitting the short (orange) and long (blue) filament populations. The total population (black) is displayed for comparison. \textit{Top left}: the GNFW model presented in Eq.~\ref{Eq:GNFW}, \textit{Top right}: the $\beta$-model of Eq.~\ref{Eq:Beta}, \textit{Bottom left}: the double power law of Eq.~\ref{Eq:2PL}, and \textit{Bottom right}: the single power law model (Eq.~\ref{Eq:1PL}).}
    \label{Fig:mcmc1}
    \end{figure*}

In Fig.~\ref{Fig:mcmc1}, we show the results of the MCMC exploration in the form of corner plots (orange is for short, blue for long, and black for all filaments). Overall, we note that the parameters are rather well constrained except for the GNFW where we observed several degeneracies. We also note that fitting short and long filaments separately yields different and distinct parameters, confirming that we are dealing with two very different populations. For illustration, we quantified these differences for the double power law model, using Eq.~\ref{Eq:delta}, and we found that the best-fit values of long filaments are 5.7, 3.3, 2.7 and 7.2 $\sigma$ away from the medians of short filaments, for the parameters $n_0$, $r_\mathrm{f}$, $\alpha$ and $\beta$ respectively.

By analysing the values of the parameters in Tables \ref{Table:fitSHORT} and \ref{Table:fitLONG}, we observe that, for all the models, short filaments have always lower $r_\mathrm{f}$ values than long filaments, meaning that the transition between the two radial regimes occurs at smaller scales for the short population. Moreover, long filaments have in general steeper slopes, especially at large radii, indicating a faster drop of the galaxy density with radial distance. This is in qualitative agreement with the findings of the previous sections.

These results can be put in context with the radial DM density profiles of filaments from N-body simulations of \cite{Colberg2005}, \cite{GonzalezPadilla2010} and \cite{AragonCalvo2010}, where it is shown that the DM density follows a power law of slope $-2$ at the outskirts of the filament. In this work, the density of galaxies around short and long filaments is found to follow outer slopes of roughly $- 1.82$ and $- 2.14$, suggesting that, at the outskirts of the filament, galaxies follow the DM  skeleton. However, for a proper comparison with the previously cited papers one should extend this work to DM density profiles of short and long filaments in order to quantify any possible bias between DM and galaxy densities.

   \begin{figure}
   \centering
   \includegraphics[width=0.5\textwidth]{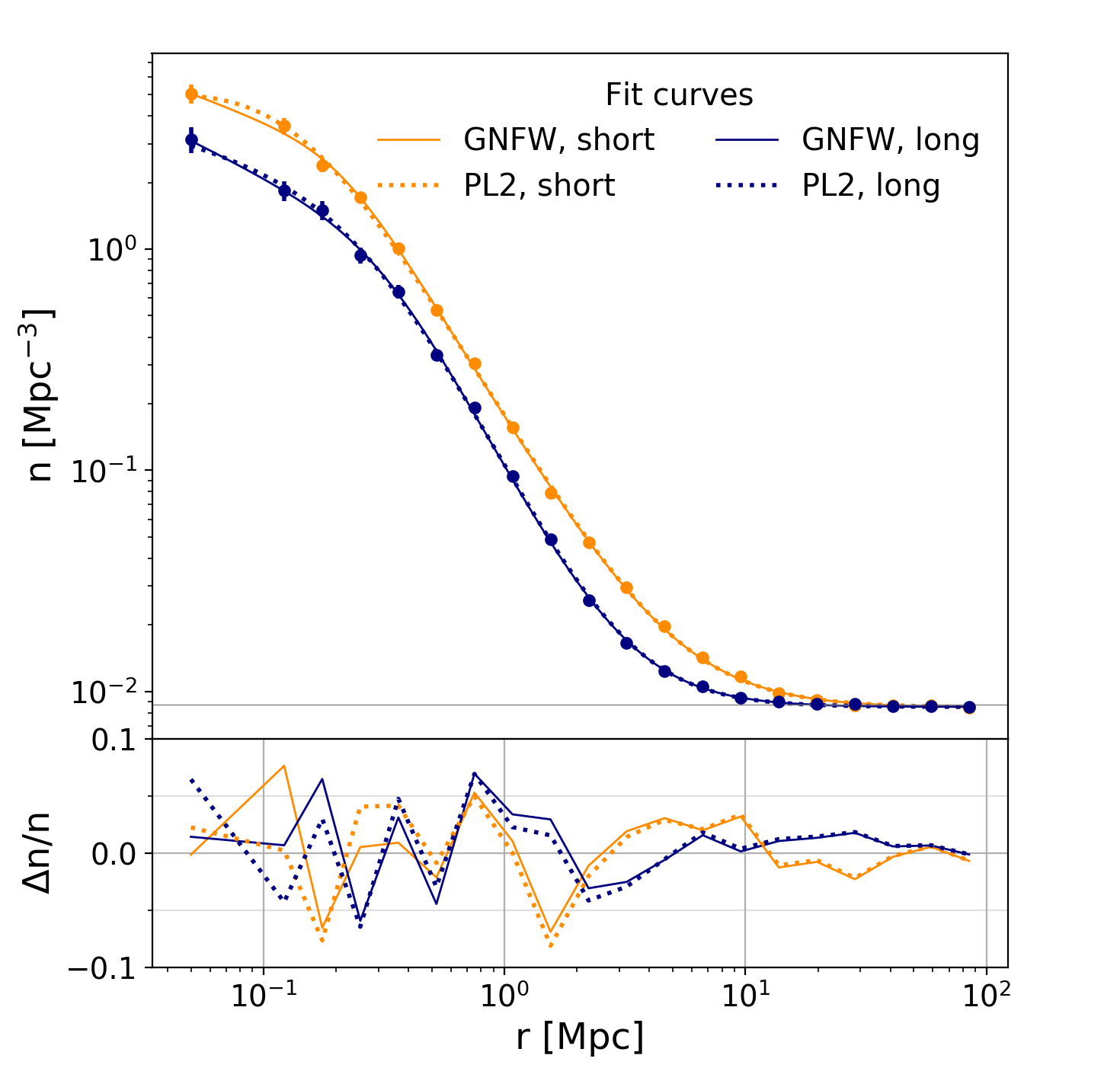}
   \caption{\textit{Top}: Fit-curves of short (orange) and long (blue) filaments. Solid and dashed lines correspond respectively to the GNFW (Eq.~\ref{Eq:GNFW}) and the double power law (Eq.~\ref{Eq:2PL}) models. The fit curves are plotted using the parameters of Tables \ref{Table:fitSHORT} and \ref{Table:fitLONG}. \textit{Bottom}: Relative difference of the fits with respect to the data, defined as: $\Delta \mathrm{n} / \mathrm{n} (r) = [\mathrm{n}^\mathrm{data}(r) - \mathrm{n}^\mathrm{model}(r)]/\mathrm{n}^\mathrm{data}(r)$.}
    \label{Fig:fit_profiles}
    \end{figure}
    
Finally, we display in the upper panel of Fig.~\ref{Fig:fit_profiles} the data points and the models obtained using the best-fit parameters of the double power law and the GNFW model plotted in dotted and solid lines, respectively. The orange colour stands for short filaments and blue is for long filaments. These two models illustrate a case where parameters are well constrained (double power law) and another with significant degeneracies (GNFW). We also show, in the lower panel of Fig.~\ref{Fig:fit_profiles}, the relative difference between the data and the model defined as $\Delta \mathrm{n} / \mathrm{n} (r) = [\mathrm{n}^\mathrm{data}(r) - \mathrm{n}^\mathrm{model}(r)]/\mathrm{n}^\mathrm{data}(r)$. We note that both models fit well the data, in agreement with the $\chi^2_\nu$ values of Fig.~\ref{Fig:chisquare}. The largest relative difference reaches $8\%$ and is observed in the inner part of the filament. Beyond $2$ Mpc from the  spine of the filaments, the relative difference is less than $4\%$ for both the GNFW and the double power law models, which additionally agree together quite well at these radii.

\section{\label{Sect:Conclusions}Conclusions}

In this paper, we analysed the filamentary structures of the cosmic web from the galaxy distributions of five different simulations. We took TNG300-1 as a reference and we compared to it the results from the Magneticum, the TNG300-2, TNG100-2 and Illustris-2 simulations. We found the following main conclusions:

   \begin{enumerate}[$\bullet$]
   \item Galaxy densities in filaments of all simulations have the highest values near the core of filaments. The density decreases with increasing radial distance. The excess of galaxy density with respect to the background density shows a significance larger than $5 \sigma$ up to 27 Mpc away from the core of the filaments in the TNG300-1 simulation.
   
   \item Baryonic physics (like feedback effects) play a role in the distribution of matter around filaments. This was clearly seen by comparing the Illustris-2 filament profiles with profiles of the other simulations. Still, as expected for dense structures, the essential and dominant driver remains gravity.

   \item In simulations with similar feedback models that have been calibrated on observations, filaments of different lengths do not have the same galaxy density profiles. In the reference TNG300-1, Kolmogorov-Smirnov tests show that short ($L_\mathrm{f} < 9$ Mpc) and long ($L_\mathrm{f} \geq 20$ Mpc) filaments constitute two different populations. A good agreement between the simulations is found in the profiles of short and long filaments (modulo resolution effects causing the enlargement of the cores).
   
    \item Short ($L_\mathrm{f} < 9$ Mpc) filaments are denser, puffier ($r_{2} \sim 5$ Mpc), statistically more connected to massive objects and likely to be found in over-dense regions. They may be interpreted as bridges of matter between over-dense structures like clusters. These results are found to be simulation-independent.
    
    \item Long ($L_\mathrm{f} \geq 20$ Mpc) filaments are less dense, thinner ($r_{2} \sim 3$ Mpc), statistically connected to less massive objects and trace under-dense environments. They may correspond to the filaments shaping the large scales of the cosmic web. These results are found to be simulation-independent.

    \item Empirical models like the GNFW profile, the $\beta$-model, the single and the double power law fit the galaxy density profiles of the two filament populations with different and distinct parameterisations for short and long filaments. The Einasto model is not adapted for the description of the galaxy density profiles of filaments.

   \end{enumerate}

These observations indicate that the short and long populations of filaments belong to different scales governed by different dynamical processes.
In over-dense regions the dynamics of matter might be dominated by gravitational forces that pull matter from the lowest to the highest density zones, whereas in under-dense regions the dark energy forces responsible for the cosmic accelerated expansion and stretching of the cosmic web might be dominant over the gravitational forces driving the collapse. 

An analysis of the velocity field around short and long filaments might be useful to unveil the dynamics and the evolution of these two different populations. Also, a complete study of the properties and the dynamics of gas around filaments will allow us to better understand and assess the effects of feedback physics on the different filament populations. Finally, the classification and characterisation of filaments within a theoretical framework \citep[e.g.][]{Feldbrugge2018} is necessary to gain a more complete picture of these different populations.

We note that this study has used a topological definition of the filaments, given by the DisPerSE algorithm. A natural follow up of this work will be to see if these conclusions stand when detecting filaments with a different algorithm (e.g. from the velocity field). Moreover, further analyses on the properties of galaxies around bridges of matter and cosmic filaments (e.g. gradients of stellar mass, star formation rate, etc) can be useful to complement the present picture.

\begin{acknowledgements}
The authors acknowledge the useful comments of the anonymous referee.
This research has been supported by the funding for the ByoPiC project from the European Research Council (ERC) under the European Union’s Horizon 2020 research and innovation program grant
agreement ERC-2015-AdG 695561. (ByoPiC, https://byopic.eu). The authors acknowledge the very useful comments and discussions with all the members of the ByoPiC team (https://byopic.eu/team/), in particular with Victor Bonjean.
We thank Klaus Dolag and Antonio Ragagnin for making the Magneticum galaxy catalogue publicly available. We also thank the Illustris-TNG team for releasing publicly the full simulated snapshots.
\end{acknowledgements}


\begin{appendix}

\section{\label{Appendix:masks} Details on the mask of nodes}

In this Appendix we give a more detailed explanation on the masking of the maximum density critical points, coinciding to the positions of the nodes, as explained in Sect.~\ref{SubSect_radial_density_profiles_METHOD}.
For the TNG300-1 simulation, in order to have an unbiased description of the density field, we use the dark matter particles to compute the radius $R_{200}$ of the spheres centered on the positions of the maximum density critical points (CPmax). We compute the $R_{200}$ radii by following an iterative scheme. For each CPmax we compute the density in spheres of increasing radii, until finding the radius for which the DM density equals $200 \, \rho_c$, where $\rho_c$ is the critical density of the Universe. 

\begin{figure}
    \centering
    \includegraphics[width=0.5\textwidth]{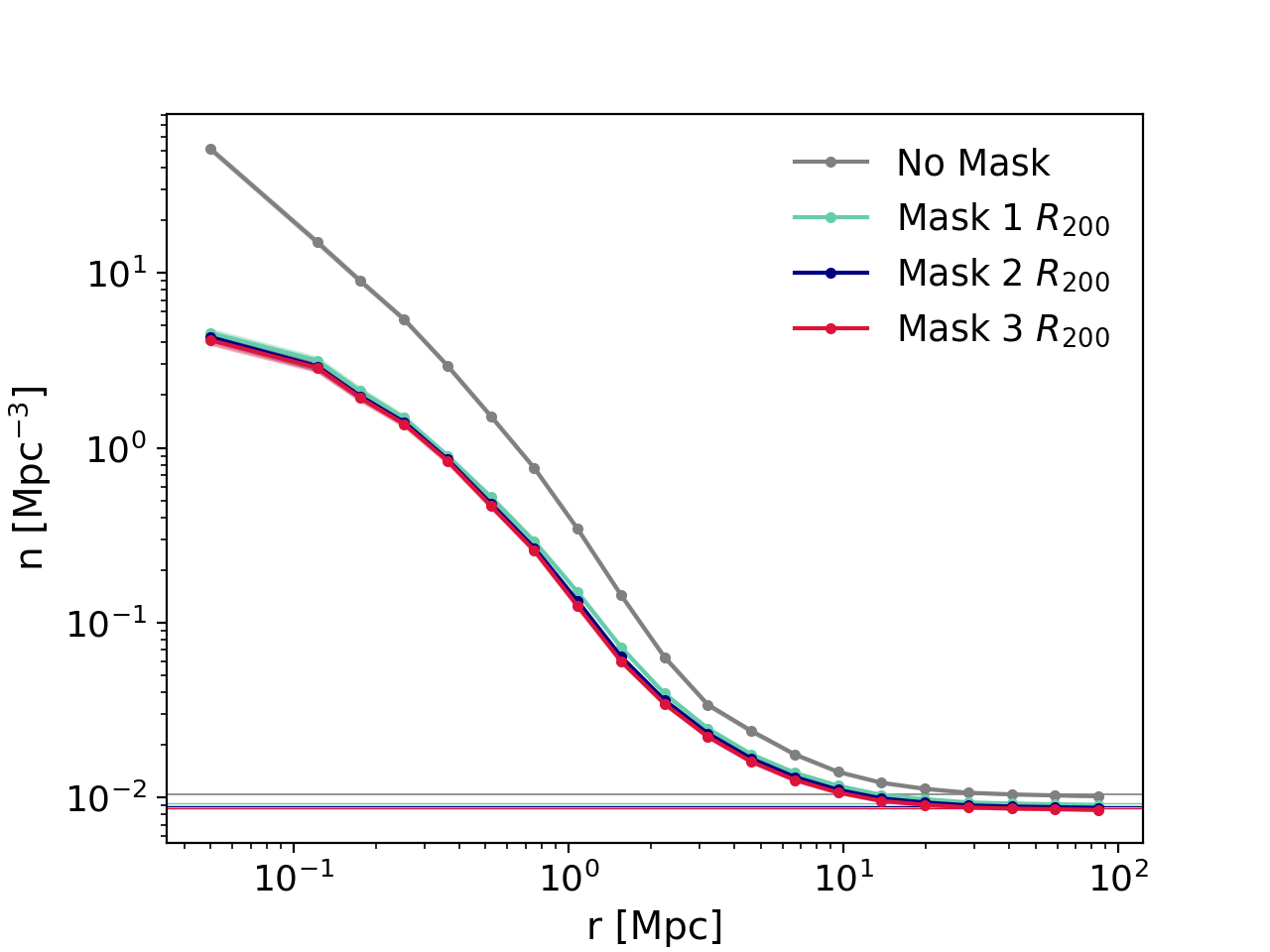}
    \caption{Density profiles of galaxies around filaments with different masks of the maximum density critical points: $1\times R_{200}$ (green), $2\times R_{200}$ (blue) and $3\times R_{200}$ (pink). The gray curve represents the density profile without applying any mask. The horizontal thin lines represent the corresponding background galaxy densities after removing the masked galaxies.}
    \label{Fig:profilesMASKS}
    \end{figure}

We compute radial density profiles (as explained in Sect.~\ref{SubSect_radial_density_profiles_METHOD}) with a mask of 1, 2 and $3 \times R_{200}$, and we compare the resulting curves with the one without any mask. The profiles are presented in Fig.~\ref{Fig:profilesMASKS}. 
We can see that, the unmasked curve (grey) is 39 $\sigma$ different from the $1 \times R_{200}$ masked profile (green), especially at the core of the filament. However, the differences between the 1, 2 or $3 \times R_{200}$ masked profiles are not significant as we get a stable signal, proving that a mask of 1, 2 or $3 \times R_{200}$ are equivalently good enough to remove the contribution from the clusters. Therefore, we make a conservative choice and adopt a $3 \times R_{200}$ mask in our analysis in order to avoid the contamination by other effects at the outskirts of clusters, like splash-back mechanisms \citep{More2015}, for example.
We also note that the density backgrounds (horizontal lines) of the masked profiles are slightly lower than the unmasked one. This is due to the fact that masked profiles account for less galaxies in the same simulation volume.

\section{\label{Appendix:AGN} Galaxy density profiles in nodes}

\begin{figure}
    \centering
    \includegraphics[width=0.5\textwidth]{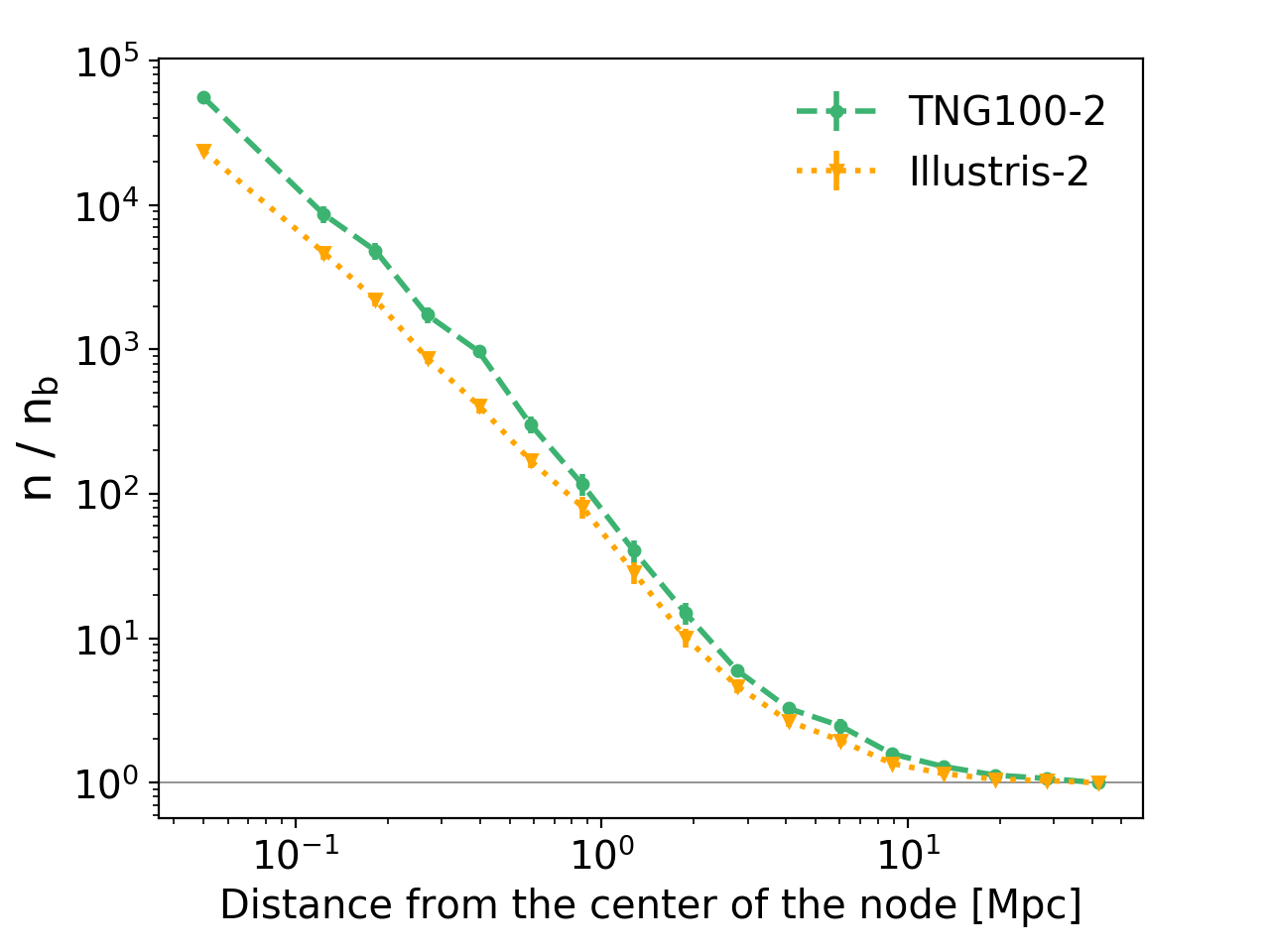}
    \caption{Galaxy density profiles in nodes, for the Illustris-2 and the TNG300-2 simulation.}
    \label{Fig:HALOprofiles}
    \end{figure}

Here we use the outputs of the Illustris-2 and TNG100-2 simulations to compute the galaxy density profiles in nodes. These are identified by the maximum density critical points of the galaxy catalogues, and masked in the analysis of filaments (see Fig.~\ref{Fig:schema_skel} and Sect.~\ref{SubSec:Disperse} and \ref{SubSect_radial_density_profiles_METHOD}). 
We assume spherical symmetry and we compute the density of galaxies in concentric spheres centered on the position of the critical point. The resulting profiles are presented in Fig.~\ref{Fig:HALOprofiles}.

The Illustris-2 profile exhibits a shallower slope and smaller values at the centre of the nodes. This was observed before by \cite{Genel2014ILLUSTRIS} and \cite{Haider2016ILLUSTRIS}, who showed that the AGN feedback model in the Illustris simulation (specially the radio-mode) is strong leading to tensions with observations. Indeed, in this simulation massive structures at low redshift are found to be almost devoid of gas as a result of the radio-mode AGN feedback \citep{Genel2014ILLUSTRIS}. This effect may also explain the shallow cores in the filament profiles of Illustris-2 simulation, presented in Fig.~\ref{Fig:profile3_nulltest}.

\end{appendix}

\bibliography{main} 

\end{document}